\documentclass[aps,pra,onecolumn,amsmath,amssymb,showpacs,amsfonts,longbibliography]{revtex4-1}
\usepackage{epsfig}
\usepackage{graphics}
\usepackage{dcolumn}
\usepackage{bm}
\usepackage{epstopdf}
\usepackage{color}
\usepackage{amsmath}
\usepackage{array}
\usepackage{natbib}
\newcolumntype{K}[1]{>{\centering\arraybackslash}p{#1}}

\begin{document}

\title{Inhomogeneous preferential concentration of inertial particles in turbulent channel flow}
\author{Lukas Schmidt$^{1}$}
\author{Itzhak Fouxon$^{2}$}
\author{Peter Ditlevsen$^{3}$}
\author{Markus Holzner$^1$}

\affiliation{$^1$ ETH Zurich, Stefano Franscini-Platz 5, 8093 Zurich, Switzerland}
\affiliation{$^2$ Department of Computational Science and Engineering, Yonsei University, Seoul 120-749, South Korea}
\affiliation{$^3$ Centre for Ice and Climate, Niels Bohr Institute, University of Copenhagen, Copenhagen, Denmark}

\begin{abstract}
Turbophoresis leading to preferential concentration of inertial particles in regions of low turbulent diffusivity is a unique feature of inhomogeneous turbulent flows, such as free shear flows or wall-bounded flows. In this work, the theory for clustering of weakly inertial particles in homogeneous turbulence of Fouxon (\emph{PRL} 108, 134502 (2012)) is extended to the inhomogeneous case of a turbulent channel flow. The inhomogeneity contributes to the cluster formation in addition to clustering in homogeneous turbulence. A space-dependent rate for the creation of inhomogeneous particle concentration is derived in terms of local statistics of turbulence. This rate is given by the sum of a fluctuating term, known for homogeneous turbulence, and the term having the form of a velocity derivative in wall-normal (y) direction of the channel. Thus particle motion can be considered as the sum of the average $y-$dependent flux to the wall (differing from average Eulerian velocity of the particles) and directionless fluctuations. The inhomogeneous flux component of the clustering rate depends linearly on the small Stokes number $St$ that measures the particle inertia. In contrast the homogeneous component has higher order smallness, scaling quadratically with $St$. We provide the formula for the pair-correlation function of concentration that factorizes in product of time and space-dependent average concentrations and time-independent factor of clustering that obeys a power-law in the distance between the points. This power-law characterizes inhomogeneous multifractality of the particle distribution. Its negative scaling exponent - the space-dependent fractal correlation codimension - is given in terms of statistics of turbulence. A unique demonstration and quantification of the combined effects of turbophoresis and fractal clustering in a direct numerical simulation of particle motion in a turbulent channel flow is performed according to the presented theory. The strongest contribution to clustering coming from the inhomogeneity of the flow occurs in the transitional region between viscous sublayer and the buffer layer. Fluctuating clustering effects as characteristic of homogeneous turbulence have the maximum intensity in the buffer layer. Further the ratio of homogeneous and inhomogeneous term depends on the wall distance. The inhomogeneous terms may significantly increase the preferential concentration of inertial particles, thus the overall degree of clustering in inhomogeneous turbulence is potentially stronger compared to particles with the same inertia in purely homogeneous turbulence. The presented theoretical predictions allow for a precise statistical description of inertial particles in all kinds of inhomogeneous turbulent flows.


\end{abstract}

\maketitle

\section{Introduction}
Particulate matter suspended in fluids is a common feature of turbulent flows encountered in industrial devices as well as the environment. Typical examples range from combustion processes in diesel engines and gas turbines \cite{Engineering2,lee2002morphological}, ash or aerosols expelled from volcanic eruptions or chemical or nuclear accidents \cite{Engineering1,hidy2012aerosols}, to liquid rain droplets in clouds \cite{FFS,shaw2003particle}. Knowledge about the distribution of these particles can be essential for the working reliability and efficiency of engines \cite{williams1979drop}, climate predictions \cite{Seinfeld,reeks2014transport} and the health of living organisms \cite{Flagan}. Therefore, understanding of particle-turbulence interaction causing heterogeneous particle distributions and identifying potential high-concentration regions is a crucial aspect. 
Clustering of inertial particles in homogeneous turbulence conditions has been subject to intensive research for a long time and is thus generally well explained by the phenomenon of small-scale or fractal clustering, see for example \cite{BecCenciniHillerbranddelta,sundaram1997collision,Stefano,JYL,BFF,CenciniBecBiferale,Cencini,monchaux2010preferential,eatonfessler}. 

Many flows in the environment or industrial applications are physically bounded by walls, which creates a heterogeneity in the flow. In these non-uniform flows, inertial particles are subject to turbophoretic forces that create a strongly inhomogeneous distribution of particles along the direction of non-uniformity. Turbophoresis is generally associated with a net flux of particles towards regions of lower turbulence diffusivity, thus leading to increased particle concentration in near-wall regions of wall-bounded flows. Since inhomogeneous turbulent flows containing non-passive particles are a common occurrence in nature and industry, turbophoresis is an ubiquitous phenomenon that deserves special attention. 
The turbophoretic mechanism was initially described from a theoretical point of view in \cite{caporaloni} and \cite{reeksturbophoresis}. The work of \cite{young} developed a theoretical description of the particle transfer to the wall caused by turbulent sweep and ejection events. Subsequently, several experimental investigations (e.g. \cite{kaftoria,kaftorib,righetti}) have confirmed the accumulation of particles in near-wall regions due to turbophoresis. Increasing computational resources allowed for detailed investigations of turbophoresis based on direct numerical simulations of inhomogeneous turbulent flows. The early work of Eaton \& Fessler \cite{eatonfessler} found heavy particles to be localized preferably in regions close to the wall that feature a low instantaneous velocity. 
Further investigations \cite{marchioli,picciotto} indicated that coherent wall structures are very likely responsible for high particle concentrations in low velocity regions in wall-proximity. Furthermore, the spatial development of particle concentrations in a turbulent pipe flow was investigated numerically in \cite{picano}. These investigations mainly focused on the increasing concentration of particles with relatively large inertia in the near-wall region and the possibility to derive information about particle surface deposition.
Preferential concentration arising due to turbophoresis as well as small-scale clustering has been investigated in a channel flow for heavy particles by \cite{sardina}. \cite{de2016clustering} studied the effects of turbophoresis and small-scale clustering in a shear-flow without walls for a wide range of particle inertia.
They showed that turbophoresis is stronger for particles with moderate inertia, whereas small-scale clustering dominates at week inertia.\\

Multifractality was derived originally for spatially uniform turbulence. It is a small-scale phenomenon holding at scales smaller than Kolmogorov scale. Recently, multifractality was generalized to the case of inhomogeneous turbulence \cite{Schmidt}. The derivation assumed that during the characteristic time of formation of fractal structures the motion of small parcels of particles is confined in a not too large region of space where statistics of flow gradients can be considered uniform. It was demonstrated that the pair-correlation function of concentration of particles factorizes in the product of (possibly time-dependent) average local concentrations and a geometrical factor of fractal increase of probability of two particles to be close. Thus we can separate dependencies in the particle distribution. The geometry of the time-dependent multifractal to which the particles are confined is statistically stationary. It is space-dependent because of the inhomogeneity of turbulence. However, the overall number of particles, that distribute over the multifractal locally, is determined by the average local concentration that can be time-dependent. For instance in the case of turbophoresis there is depletion of local average concentration because of the particle flux to the wall.

The present work provides an extensive theoretical and numerical investigation of the clustering degree of weakly inertial particles in a turbulent channel flow. The specific case of interest here is the turbophoretic behavior of weakly inertial particles in flows where the turbulent diffusivity varies in one direction (i.e. wall-normal) but the turbulence is homogenous in wall-parallel planes. The universal framework of weakly compressible flow \cite{FFS,fouxon1,fphl} is therefore modified to describe the combined clustering effects occurring due to the inhomogeneity of the flow and fluctuational clustering present in homogeneous turbulence. The theory is used as basis for the analysis of inertial particles in a direct numerical simulation (DNS) of a turbulent channel flow.

Our main theoretical result in this paper is the formula for the space-dependent rate $\left\langle \sum\lambda_i(y)\right\rangle$ of creation of inhomogeneities of concentration of particles in terms of local statistics of turbulence, 
\begin{eqnarray}&&\!\!\!\!\!\!\!\!
\left\langle \sum\lambda_i(y)\right\rangle=\!-\tau^2\int_0^{\infty}\langle\nabla^2p(0)\nabla^2p(t)\rangle_c dt'+\partial_y v_{eff}(y),\ \ \ \ v_{eff}(y)=-\tau\partial_y  \langle u_y^2\rangle+\tau\int_0^{\infty}\langle u_y(0) \nabla^2p(t)\rangle_c dt. \label{sum}
\end{eqnarray}
In this formula $y$ is the distance to the wall, $\tau$ is the Stokes time of the particles, $\bm u$ and $p$ are the turbulent flow velocity and pressure, respectively. The angular brackets designate averaging over local statistics of turbulence and brackets with $c$ standing for the dispersion. Besides the average Eulerian velocity of the particles $-\tau\partial_y  \langle u_y^2\rangle$, the rest of the terms come from distinction between Lagrangian and Eulerian averages in the formation of particle inhomogeneities. The first term is the local form of the formula known for spatially uniform turbulence that holds in the bulk. Inhomogeneity of turbulence produces an effective velocity term $v_{eff}(y)$ that in contrast with the bulk term is proportional to the first power of $\tau$. Thus clustering of weakly inertial particles can be much stronger because of inhomogeneity effects. 

The numerical investigation of this work is based on DNS data obtained from the Johns Hopkins University Turbulence Database (JHTDB). We implement a Lagrangian tracking algorithm for inertial particles based on time-resolved Eulerian DNS results of a turbulent channel flow. This allows to evaluate a wide range of flow quantities along particle trajectories which serves as basis for the quantification of preferential concentration of inertial particles in inhomogeneous turbulence. 

This paper commences with a theoretical analysis (Sec.\,\ref{sec:theory}) and the equations used for the analysis of the numerical data. This is followed by a description of the channel flow simulation details, including the implementation of the particle tracking algorithm and a brief characterization of the flow quantities of interest (Sec.\,\ref{sec:dns}). Subsequently, the performed analysis and results based on the direct numerical simulation (Sec.\,\ref{sec:analysis}) are illustrated. A summary of these results and the corresponding conclusions are presented in Sec.\,\ref{sec:discussion}.

\section{Sum of Lyapunov exponents}
\label{sec:theory}
In this Section, we introduce the local rate of production of inhomogeneities of concentration of particles. This rate is the sum of Lyapunov exponents $\sum\lambda_i$. We observe that the local rate of particle density increase is determined by the local divergence of velocity. However that divergence is fluctuating in turbulence. The sum of Lyapunov exponents describes the rate of accumulated growth obtained by proper averaging over the fluctuations and it differs both from fluctuating and average divergences. We will see in the coming Sections that it is the average divergence plus the coherent contribution of the fluctuations.       

We consider particles with weak inertia suspended in incompressible turbulent flow in a channel. The strength of inertia of a small spherical particle is quantified by the particle Stokes time $\tau$. This is given by $\tau=2a^2\rho_p/(9\nu\rho)$, with $a$ being the particle radius, $\rho$ and $\rho_p$ as the fluid and particle density respectively and $\nu$ is the fluid viscosity. We consider the case of dense particles with $\rho_p\gg \rho$ so effects such as memory and added mass can be neglected \cite{MaxeyRiley}. Assuming that the Reynolds number $Re_p$ of the flow perturbation caused by the particles is small ($Re_p\ll 1$) we can use the linear law of friction finding the equation of motion,
\begin{equation}
\frac{d\bm x}{dt}=\bm v,\ \ \frac{d\bm v}{dt}=-\frac{\bm v-\bm u[t, \bm x(t)]}{\tau}.\label{lw}
\end{equation}
Here, $\bm x(t)$ is the particle coordinate, $\bm v(t)$ is the velocity and $\bm u(t, \bm x)$ is the turbulent flow velocity. Since we consider particles with small inertia, the velocity may be approximated as \cite{Maxey},
\begin{equation}
\bm v\,=\,\bm u-\tau[\partial \bm u/\partial t + (\bm u\cdot \bm\nabla)\bm u] ,
\label{eq:particlevelocity}
\end{equation} where $\tau[\partial \bm u/\partial t + (\bm u\cdot \bm\nabla)\bm u]$ represents the inertial particle drift. 


Despite that the fluid velocity is divergence-free, the inertial drift of the particle velocity results in a 'weakly compressible' particle flow because $\nabla\cdot\bm v\neq 0$. For the case where particles are seeded into a statistically stationary, incompressible channel flow, with two homogeneous $(x,z)$ and one inhomogeneous ($y$) flow direction the mean Eulerian divergence of the particle velocity field reduces to,
\begin{equation}
 \langle v_y\rangle=-\tau\langle \nabla \cdot (\bm u u_y)\rangle=-\tau\partial_y  \langle u_y^2\rangle\ \ \langle \nabla\cdot\bm v\rangle=\frac{\partial \langle v_y\rangle}{\partial y}=-\tau\frac{\partial^2 \langle u_y^2\rangle}{\partial y^2}.
\label{eq:divv_inh}
\end{equation}
The formula for $\langle v_y\rangle$, obtained by averaging Eq.~(\ref{eq:particlevelocity}), is the well-known result for the average Eulerian particle velocity towards the wall. Since $\langle u_y^2\rangle$ monotonously grows away from the wall reaching a maximum in the log-law region of the channel, then $\langle v_y\rangle$ is directed toward the walls constituting the simplest demonstration of turbophoretic motion of particles towards the regions with smaller intensity of turbulence. In contrast, $\langle \nabla\cdot\bm v\rangle$ is not sign-definite: near the wall where $\langle u_y^2\rangle$ has minimum we have positive $\partial_y^2 \langle u_y^2\rangle$ (right on the wall $\partial_y^2 \langle u_y^2\rangle=0$ but it is positive nearby) and negative $\langle \nabla\cdot\bm v\rangle$, conversely we have $\partial_y^2 \langle u_y^2\rangle<0$ and $\langle \nabla\cdot\bm v\rangle>0$ already in the buffer layer. Thus there is a point $y_*$ where $\langle \nabla\cdot\bm v\rangle=0$ so $\langle \nabla\cdot\bm v\rangle$ is negative for $y<y_*$ and positive otherwise.

Based on Eulerian divergence we could conclude that at $y<y_*$ there are positive correlations in positions of particles separated by distance smaller than the viscous scale $\eta$. These particles are in the same divergence \cite{frisch} that is negative on average and thus approach each other. In contrast, for $y>y_*$ there would seem to be negative correlations of positions. In reality the particles move and correlate in reaction not to Eulerian divergence of the flow but to the divergence of the flow in the frame of reference that moves with the particle. This divergence determines the evolution of infinitesimal volumes $V$ of particles that obey \cite{Batchelor}, 
\begin{equation}
\frac{d\ln V}{dt}=w(t, \bm x(t)),\ \ w(t, \bm x)=\nabla\cdot\bm v(t, \bm x),
\end{equation}
where $\bm x(t)$ is the trajectory of some particle located inside the considered infinitesimal volume. This equation holds provided the largest linear size of $V$ is much smaller than $\eta$. The solution of this equation for the logarithmic increase rate of $V(t)$ defines the finite-time sum of Lyapunov exponents $\sum\lambda_i(t)$ as,
\begin{equation}
\sum_{i=1}^3\lambda_i(t)=\frac{1}{t}\ln \left(\frac{V(t)}{V(0)}\right)=\frac{1}{t}\int_0^t w(t', \bm x(t'))dt'. \label{finite}
\end{equation}
The RHS of this equation has the form that occurs in the ergodic theorem \cite{Sinai}. Indeed, in spatially uniform turbulence ergodicity implies that the RHS converges to the deterministic limit in the limit of $t\to\infty$ (we remark that the convergence holds for almost every point with respect to the stationary measure but in this context it holds for almost every point in space that is with possible exception of points with zero total volume). This limit is independent of the initial position of $V(t)$. Thus the limit equals to the average over the initial position defining the sum of Lyapunov exponents of spatially uniform turbulence \cite{review,ff},
\begin{equation}
\left(\sum \lambda_i\right)_{bulk}=\lim_{t\to\infty}\sum\lambda_i(t)=\lim_{t\to\infty}\frac{1}{t}\int_0^t \langle w(t', \bm x(t'))\rangle dt'=\lim_{t\to\infty}\langle w(t, \bm x(t))\rangle, 
\end{equation}
where angular brackets stand for averaging over $\bm x(0)$. Here the subscript signifies that this formula holds in the bulk of the flow, far from the boundaries, where the spatially uniform turbulence framework applies. It was demonstrated in \cite{ff}, and will be proved below somewhat differently, that this can be written as,
\begin{equation}
\left(\sum \lambda_i\right)_{bulk}=-\int_0^\infty \langle w(0)w(t)\rangle dt, \label{fluct}
\end{equation}
which clarifies that $\sum\lambda_i<0$ see details in \cite{ff}. Taking the divergence of the Navier-Stokes equations obeyed by $\bm u$ we find that $w=\tau\nabla^2 p$ where $p$ is the turbulent pressure. Thus we can write
\begin{equation}
\left(\sum \lambda_i\right)_{bulk}=-\tau^2 \int_0^\infty \langle\nabla^2p(0)\nabla^2p(t)\rangle dt\propto St^2, \label{blk}
\end{equation}
where we defined the Stokes number $St=\tau\sqrt{\epsilon/\nu}$. Here $\epsilon$ is the average rate of energy dissipation per unit mass.

In practice, the infinite time limit must be understood as near convergence beyond a certain convergence time $t_c$ when for most of the trajectories $\bm x(t)$, the LHS of Eq.~(\ref{finite}) becomes constant at $t\sim t_{c}$. This time is strongly different for $\sum\lambda_i$ and other combinations of Lyapunov exponents. These are defined very similarly where $\lambda_1$ and $\lambda_1+\lambda_2$ are logarithmic increase rates of infinitesimal line and surface elements, respectively \cite{review}. For instance 
\begin{equation}
\lambda_1(t)+\lambda_2(t)=\frac{1}{t}\ln \left(\frac{S(t)}{S(0)}\right), \label{area}
\end{equation}
where $S(t)$ is the infinitesimal area of fluid particles \cite{review,Schmidt}. For $\lambda_1$ and $\lambda_2$ the convergence time $t_c$ is a few Kolmogorov times $\tau_{\eta}$. Here $\tau_{\eta}=\sqrt{\nu/\epsilon}$, which is usually defined as typical time-scale of turbulent eddies at the viscous scale \cite{frisch}, is also the correlation time of flow gradients (and thus $w$) in the fluid particle frame \cite{review,ff}. In contrast, for $\sum\lambda_i$ the convergence time is much longer as seen considering the dispersion the of $\sum\lambda_i(t)$, 
\begin{eqnarray}&&\!\!\!\!\!\!\!\!\!\!\!\!\!\!
\left\langle \left(\sum\lambda_i(t)\right)^2\right\rangle\!-\!\left\langle \sum\lambda_i(t)\right\rangle^2\!=\!\frac{1}{t^2}\int_0^t \!\!dt_1dt_2  \langle w(t_1, \bm x(t_1))w(t_2, \bm x(t_2))\rangle_c\!\sim \!\frac{1}{t}\int_0^\infty \!\!\langle w(0)w(t)\rangle dt\sim \frac{|\sum\lambda_i|}{t},\ \ t\gg \tau_{\eta}, \label{sps}
\end{eqnarray}
where $c$ stands for cumulant (dispersion) and we used that the correlation time $\tau_{\eta}$ of $w(t, \bm x(t))$ is much smaller than $t$. We observe that the normalized rms deviation of $\sum\lambda_i(t)$ obeys,
\begin{equation} \frac{\sqrt{\left\langle \left(\sum\lambda_i(t)\right)^2\right\rangle-\left\langle \sum\lambda_i(t)\right\rangle^2}}{|\sum\lambda_i(t)|}\sim \frac{1}{\sqrt{|\sum\lambda_i| t}}\sim \frac{1}{St \sqrt{t/\tau_{\eta}}}.\label{rms}
\end{equation}
Thus fluctuations of $\sum \lambda_i(t)$ are small only at quite large times $t\gg \tau_{\eta}/St^2$. This is because non-zero $\sum\lambda_i$ is due to fluctuations, as evident from Eq.~(\ref{fluct}), so dispersion of $\sum \lambda_i(t)$ has the same order in $St$ as the average.  Thus the convergence time for the sum of Lyapunov exponents obeys $t_c\sim \tau_{\eta}/St^2$. This slow convergence of long-time limit $\sum\lambda_i(t)$ at small $St$, seems to be unobserved previously.

Clearly the long-time convergence of $t^{-1}\ln V(t)/V(0)$ to a deterministic limit makes it a very useful quantity because it allows a deterministic prediction despite the randomness of turbulence. We determine $\sum\lambda_i$ once and then we can predict it for any arbitrary trajectory. The sum of the Lyapunov exponents, made dimensionless by dividing by a factor of order of $\sqrt{\epsilon/\nu}$ determines the strength of the clustering by giving fractal dimensions of the random attractor formed by particles in space, see e. g. \cite{fouxon1} and Section \ref{pc}. 

Fortunately, we can extend the notion of the sum of Lyapunov exponents to the inhomogeneous case with which we can describe clustering closer to the walls. We consider the situation when $\bm x(t_c)$ does not deviate from $\bm x(0)$ by the characteristic scale of inhomogeneity of turbulent statistics which is the distance to the wall $y$. This demands that the $y-$dependent inhomogeneity time $t_{in}(y)=y/\langle u_y^2\rangle^{1/2}$, where $\langle u_y^2\rangle^{1/2}$ is the typical transversal velocity, is much larger than $t_c$. Here $t_{in}(y)$ is the time during which the trajectory passes a distance comparable with $y$ over which the statistics changes. If $t_c\ll t_{in}(y)$ holds then during the convergence time $\bm x(t)$ in Eq.~(\ref{finite}) stays in the region where the statistics is roughly uniform. Thus we can use the results for spatially uniform turbulence locally. We could then guess that, 
\begin{equation}
\left(\sum \lambda_i\right)_{bulk}(y)\sim -\tau^2 \int_0^\infty \langle\nabla^2p(0, y)\nabla^2p(t)\rangle dt, \label{sps2}
\end{equation}
where the RHS depends on the $y$ coordinate. Here the local correlation function can be defined with the help of temporal averaging. In the case of the channel instead of time averaging we can use averaging over the symmetry plane of the statistics,
\begin{eqnarray}&&\!\!\!\!\!\!\!\!\!\!\!\!\!\langle\nabla^2p(0, y)\nabla^2p(t)\rangle=\!\!\lim_{S\to\infty}\!\int \!\!\frac{dx dz}{S} \nabla^2p(0, x, y, z))\nabla^2p[t, \bm q(t, x, y, z)],\end{eqnarray} where $S$ is the area of the $x-z$ plane and we introduced 'Lagrangian' trajectories of particles labeled by their initial positions at $t=t_0$,
\begin{eqnarray}&&
\partial_t \bm q(t| t_0, \bm x)=\bm v[t, \bm q(t| t_0, \bm x)],\ \ \bm q(t| t_0, \bm x)=\bm x,\ \ \bm q(t, \bm x)=\bm q(t| t_0=0, \bm x). \label{bas2}
\end{eqnarray}
The temporal and "planar" averages are identical because the temporal average \\$\langle\nabla^2p(0, y)\nabla^2p(t)\rangle_t$ is independent of $x$ and $z$ so that,
\begin{equation}
\langle\nabla^2p(0, y)\nabla^2p(t)\rangle_t=\lim_{S\to\infty}\!\int \!\!\frac{dx dz}{S} \langle\nabla^2p(0)\nabla^2p(t)\rangle_t 
\end{equation} where interchanging the order of averages in the last term proves that time and plane averages coincide (time average of the plane average is the plane average). Below we will use planar averaging designating it by angular brackets and demonstrate that the guess given by Eq.~(\ref{sps2}) is incomplete.

\section{Identity for Lagrangian averages}

In this Section we derive an identity for Lagrangian average $\langle f(t, y)\rangle$ of an arbitrary stationary random function $f(t, \bm x)$ in the frame of particles released at the same distance $y$ from the wall,
\begin{equation}
\langle f(t, y)\rangle=\int\frac{dxdz}{S} f(t, \bm q(t, x, y, z)).
\end{equation}
Here this quantity is of interest in the case of $f=w(t, \bm x)$ when it provides the average sum of Lyapunov exponents studied in the next Section. However other cases of this quantity can be of interest in future studies so we keep arbitrariness of $f$. The derivation is a changed line of thought that appeared in \cite{ff}. We observe that $\bm q(t| t_0, \bm x)$  as a function of the initial time $t_0$ obeys,
\begin{eqnarray}&&
\partial_{t_0} \bm q(t| t_0, \bm x)+[\bm v(t_0, \bm x)\cdot\nabla]\bm q(t| t_0, \bm x)=0.\label{initime}
\end{eqnarray}
This expresses that changing initial time and position so that the initial position stays on the same trajectory does not change that trajectory:
$\bm q(t| t_0+\epsilon, \bm q(t_0+\epsilon|t_0, \bm r))=\bm q(t|t_0, \bm r)$ (the trajectory that passes through $\bm r$ at time $t_0$ is the same trajectory that passes through $\bm q(t_0+\epsilon|t_0, \bm r)$ at time $t_0+\epsilon$). Differentiating over $\epsilon$ and setting $\epsilon=0$ one finds the equation above.

We introduce the two-time version of average of $w$,
\begin{eqnarray}&&\!\!\!\!\!\!\!\!\!\!\!\!\!
\langle f(t, t_0, y)\rangle=\lim_{S\to\infty}\int \frac{dx dz}{S} f[t, \bm q(t| t_0, x, y, z)].\label{def}
\end{eqnarray}
Because of stationarity $\langle f(t, t_0, y)\rangle$ depends on $t $, $t_0$ only through the difference of the time arguments $t-t_0$. We consider the time derivative of Eq.~(\ref{def}) over $t_0$ using Eq.~(\ref{initime}),
\begin{eqnarray}&&\!\!\!\!\!\!\!\!\!\!\!\!\!
\frac{\partial \langle f\rangle}{\partial t_0 }\!=\!-\!\lim_{S\to\infty}\int \!\!\frac{dx dz}{S} v_i(t_0, x, y, z)\nabla_i f[t, \bm q(t| t_0, x, y, z)]\nonumber\\&& \!\!\!\!\!\!\!\!\!\!\!\!\!=\!-\!\lim_{S\to\infty}\int \!\!\frac{dx dz}{S} \nabla_i \left(v_i(t_0, x, y, z)f[t, \bm q(t| t_0, x, y, z)]
\right)
\nonumber\\&& \!\!\!\!\!\!\!\!\!\!\!\!\!+\langle w(0)f(t-t_0)\rangle,
\label{der1}
\end{eqnarray}
where we used stationarity and defined the correlation function of $w$,
\begin{eqnarray}&&\!\!\!\!\!\!\!\!\!\!\!\!\!
\langle w(0)f(t)\rangle\!=\!\!\lim_{S\to\infty}\!\int \!\!\frac{dx dz}{S} w(0, x, y, z))f[t, \bm q(t, x, y, z)].
\end{eqnarray}
Finally, observing that derivatives over $x$ and $z$ in Eq.~(\ref{der1}) give zero as integrals of complete derivative and taking the $y-$derivative outside the integration we find 
\begin{eqnarray}&&\!\!\!\!\!\!\!\!\!\!\!\!\!\!\!\!
\frac{\partial \langle f(t, t_0, y)\rangle}{\partial t_0}\!=\!\langle w(0)f(t-t_0)\rangle\!-\!\partial_y\langle v_y(0)f(t-t_0)\rangle,\label{der}
\end{eqnarray}
where we defined the correlation function,
\begin{eqnarray}&& \!\!\!\!\!\!\!\!\!\!\!\!\!
\langle v_y(0)f(t)\rangle\!=\!\!\lim_{S\to\infty}\!\int \!\!\frac{dx dz}{S} v_y(0, x, y, z))f[t, \bm q(t, x, y, z)].
\end{eqnarray}
We find integrating Eq.~(\ref{der}) over $t_0$ from $t_0=0$ up to $t_0=t$ and using $\langle f(t, y)\rangle=\langle f(t, t_0=0, y)\rangle$ that 
\begin{eqnarray}&&\!\!\!\!\!\!\!\!\!\!\!\!\!
\langle f(t, y)\rangle\!-\langle f(y)\rangle=\!\int_0^t \left[\partial_y\langle v_y(0)f(t')\rangle\!-\!\langle w(0)f(t')\rangle\right]dt'.\label{integral}
\end{eqnarray}
Here $\langle f(y)\rangle=\langle f(t=0, y)\rangle$ is the Eulerian average of $f(\bm x)$ over the horizontal plane. The RHS describes the difference between Lagrangian and Eulerian averages that holds because of preferential concentration of particles and thus quantifies the strength of the clustering. This integrals in Eq.~(\ref{integral}) do not necessarily converge in the long-time limit. We separate the possibly divergent term introducing cumulants (dispersion)
\begin{eqnarray}&&\!\!\!\!\!\!\!\!\!\!\!\!\!
\langle v_y(0)f(t')\rangle=\langle v_y(0)f(t')\rangle_c+\langle v_y(0)\rangle\langle  f(t')\rangle,\nonumber\\&& \!\!\!\!\!\!\!\!\!\!\!\!\!
\langle w(0)f(t')\rangle=\langle w(0)f(t')\rangle_c+\langle w(0)\rangle\langle f(t')\rangle,
\label{eq:cumulants}
\end{eqnarray}
where the angular brackets stand for average over $x-z$ coordinates. We find using $\partial_y \langle v_y(0)\rangle=\langle w(0)\rangle$ that 
\begin{eqnarray}&&\!\!\!\!\!\!\!\!\!\!\!\!\!
\langle f(t, y)\rangle\!=\langle f(y)\rangle+\int_0^t \left[\partial_y\langle v_y(0)f(t')\rangle_c-\langle w(0)f(t')\rangle_c\right]dt'-\tau(\partial_y  \langle u_y^2\rangle)\partial_y \int_0^t \langle  f(t')\rangle dt'.\label{identity}
\end{eqnarray}
where we used $\langle v_y(0)\rangle$ provided by Eq.~(\ref{eq:divv_inh}). The last term does not necessarily converge in $t\to\infty$ limit. For instance if $\langle f(t)\rangle$ has a finite long-time limit then this term grows linearly with time. We consider this identity in the case of our interest.

\section{Space-dependent sum of Lyapunov exponents for channel turbulence} 
\label{sec:lyapunov}

In this Section we study the average sum of Lyapunov exponents. That describes the average logarithmic rate of growth of infinitesimal volumes that start at the same distance from the wall $y$. Performing averaging over initial $x-z$ coordinates of the volume we find from Eq.~(\ref{finite}),  
\begin{equation}
\left\langle \sum\lambda_i(t, y)\right\rangle=\frac{1}{t}\int_0^t \langle w(t', y)\rangle dt'=\frac{1}{t}\int_0^t dt'\int\frac{dxdz}{S} w(t', \bm q(t', x, y, z)). \label{dfnt}
\end{equation}
The identity given by Eq.~(\ref{identity}) gives in the leading order in $St$ for $f=w$,
\begin{eqnarray}&&\!\!\!\!\!\!\!\!\!\!\!\!\!
\langle w(t, y)\rangle\!=\tau\partial_y\int_0^t \langle u_y(0)\nabla^2 p(t')\rangle_c dt'-\tau^2\int_0^t \langle \nabla^2 p(0)\nabla^2 p(t')\rangle_c dt'-\tau^2(\partial_y  \langle u_y^2\rangle)\partial_y \int_0^t \langle  \nabla^2 p(t')\rangle dt'-\tau\partial_y^2  \langle u_y^2\rangle.\label{complete}
\end{eqnarray}
where we used $\langle w\rangle$ given by Eq.~(\ref{eq:divv_inh}). In different time correlation functions in this formula we can use trajectories of the tracers in the leading order in $St$. The first term in the integrand is of order $St$ while the second is of order $(St)^2$. However both terms have to be kept because the former vanishes in the bulk due to spatial uniformity. In contrast, the third term in the integrand which is of order $(St)^2$ is smaller than the first term and can be neglected. The remaining integrals have finite $t\to\infty$ limit. We make the  plausible assumption that the integrals converge over the local Kolmogorov time-scale $\tau_{\eta}(y)=\sqrt{\nu/\epsilon(y)}$. Then we find that,
\begin{eqnarray}&&\!\!\!\!\!\!\!\! \langle w(t, y)\rangle=\!-\tau^2\int_0^{\infty}\langle\nabla^2p(0)\nabla^2p(t)\rangle_c dt'+\partial_y v_{eff}(y),\ \ v_{eff}(y)=\langle v(y)\rangle+\int_0^{\infty}\langle u_y(0) w(t)\rangle_c dt,\ \ t\gg \tau_{\eta}(y).\label{eq:29}\end{eqnarray}
Finally, we find Eq.~(\ref{sum}) using Eq.~(\ref{dfnt}). This formula is valid at not too large times because the third term in Eq.~(\ref{complete}) grows with time with $\tau\langle \nabla^2 p\rangle\approx \sum\lambda_i=const$ in the integrand. This term becomes non-negligible at times $t$ obeying $\tau t\partial_y  \langle u_y^2\rangle \partial_y \ln |\sum\lambda_i(y)|\gtrsim 1$. Since the scale of variations of the involved quantities is $y$ then this gives equality $\tau t  \langle (u_y/y)^2\rangle\gtrsim 1$. For spatially uniform turbulence, applicable in the bulk, using Kolmogorov theory we find that $\langle (u_y/y)^2\rangle\sim \epsilon^{2/3}y^{-4/3}$ is smaller than $\langle (\nabla u)^2\rangle$ by a factor of the Reynolds number $Re$. This would give $t\gtrsim (Re/St)\tau_{\eta}$. If we assume that this time is much larger than $t_c$ which demands $Re\gg St^{-1}$ then Eq.~(\ref{sum}) describes $\sum\lambda_i(t)$ for one trajectory so we can use $\sum\lambda_i(t)\approx \langle\sum\lambda_i(t)\rangle$. A similar consideration can be made in the near wall region of small $y$ where $\langle u_y^2\rangle\propto y^4$. Below we assume that the correction term is negligible in regions of interest (as confirmed by results of a DNS in section \ref{sec:analysis}). 

The formula given by Eq.~(\ref{sum}) has reductions in the bulk and in the turbulent boundary layer. In the bulk the statistics is uniform and we recover Eq.~(\ref{blk}) where there is no dispersion sign in the average because $\langle \nabla^2p \rangle=0$ in the bulk. In spatially uniform turbulence the sum of Lyapunov exponents is of order $St^2$. In contrast, the inhomogeneous terms are proportional to $St$ and dominate regions of strong inhomogeneity,
\begin{eqnarray}&&\!\!\!\!\!\!\!\!
\left(\left\langle\sum \lambda_i\right\rangle\right)_{inhom}\approx \partial_y v_{eff}\label{eq:30}\propto St.
\end{eqnarray}
We see that the RHS has the form of the divergence of an effective velocity which is the average Eulerian velocity of particles plus the correction. We can interpret $v_{eff}$ as the average velocity of turbophoretic particles to the wall which differs from the average Eulerian velocity $\langle v_y\rangle$ because of the difference between Lagrangian and Eulerian averages. The reason $\left(\left\langle\sum \lambda_i\right\rangle\right)_{inhom}$ scales linearly with $St$ is that in the bulk non-zero $\sum\lambda_i$ appears because of fluctuations and thus is proportional to $St^2$ but in the boundary layer there is an average effect proportional to $St$.

The complete equation (\ref{complete}) is an identity that holds for arbitrary $y$ including those close to the wall. In the passages between this identity and Eq.~(\ref{sum}) we introduced the assumption that the third term in Eq.~(\ref{complete}) can be neglected and the integrals converge over the local Kolmogorov time-scale $\tau_{\eta}(y)$. In the case of inhomogeneous statistics the integrands can have non-trivial time-dependence because the trajectory samples regions with statistics different from that at the initial point. Nevertheless the convergence seems reasonable. 

We consider the question of how well $\left\langle \sum\lambda_i(y)\right\rangle$ approximates the fluctuating finite-time Lyapunov exponent $\sum \lambda_i(t)=t^{-1}\ln V(t)/V(0)$ for volumes whose initial vertical position is $y$ and horizontal position is arbitrary. Proceeding as we did in studying the similar question in the spatially uniform situation, see Eq.~(\ref{sps2}), we consider the dispersion at $t\gg \tau_{\eta}$, 
\begin{eqnarray}&&\!\!\!\!\!\!\!\!\!\!\!\!\!\!
\left\langle \left(\sum\lambda_i(t)\right)^2\right\rangle\!-\!\left\langle \sum\lambda_i(t)\right\rangle^2\!=\!\frac{1}{t^2}\int_0^t \!\!dt_1dt_2  \langle w(t_1, \bm x(t_1))w(t_2, \bm x(t_2))\rangle_c\!\sim \!\frac{1}{t}\int_0^\infty \!\!\langle w(0)w(t)\rangle_c,\end{eqnarray}
where we used Eq.~(\ref{finite}) and consider $t\ll t_{in}(y)$ so the statistics of $w(t)$ does not change over the considered time interval. The difference from the spatially uniform case is that the last integral is no longer necessarily $\sum\lambda_i$, see Eq.~(\ref{sum}). For $y$ which are not too far from the bulk so we have $\tau^2\int_0^{\infty}\langle\nabla^2p(0)\nabla^2p(t)\rangle_c dt'\gtrsim |\partial_y v_{eff}(y)|$, see Eq.~(\ref{sum}), we can use Eq.~(\ref{rms}). Thus for these $y$ we can use for one trajectory $\sum\lambda_i(t)\approx \langle\sum\lambda_i(t)\rangle$ for $t\gtrsim \tau_{\eta}(y)/St^2(y)$ where $St(y)=\tau/\tau_{\eta}(y)$. In contrast, when $y$ is closer to the wall region where $\tau^2\int_0^{\infty}\langle\nabla^2p(0)\nabla^2p(t)\rangle_c dt'\ll |\partial_y v_{eff}(y)|$ we have in that region, 
\begin{equation} \frac{\sqrt{\left\langle \left(\sum\lambda_i(t)\right)^2\right\rangle-\left\langle \sum\lambda_i(t)\right\rangle^2}}{|\sum\lambda_i(t)|}\sim \frac{\sqrt{\int_0^\infty \!\!\langle w(0)w(t)\rangle_c}}{|\sum\lambda_i|\sqrt{ t}},
\end{equation}
cf. Eq.~(\ref{rms}). In this case convergence is faster. 

\section{Preferential concentration and distinction between Kaplan-Yorke and correlation dimensions}\label{pc}
The theory described in previous Sections provided the local rate of production of inhomogeneities of inertial particles in channel turbulence. This rate provides the growth of concentration of particles $n(t, \bm x)$. Solving the continuity equation at $t>0$, we have
\begin{eqnarray}&& 
\frac{1}{t}\ln \frac{n(t, \bm q(t, \bm x))}{n(0, \bm x)}=-\frac{1}{t}\int_0^t w(t', \bm q(t', \bm x))dt',\label{slv}
\end{eqnarray}
where the RHS is $\sum\lambda_i$ up to the sign. This formula is implied by mass conservation in the particle's frame, $n(t)V(t)=const$. We see from Eq.~(\ref{slv}) that concentration grows at large times as $\exp\left[|\langle\sum\lambda_i\rangle|t\right]$. We consider the history of creation of fluctuation of concentration at scale $r\ll \eta$, see  \cite{FFS,fphl}. This starts from compression of the volume of particles whose initial size is the correlation length $\eta$ of $w$. Over the initial volume the concentration is effectively uniform \cite{fouxon1}. The smallest dimension of the compressed volume decreases with time $t$ as $\eta\exp[-|\lambda_3|t]$ where $\lambda_3$ is the third Lyapunov exponent \cite{Schmidt,review}. Since $\lambda_3$ is non-zero for tracers then, in the leading order in Stokes number, we can use $\lambda_3$ of the fluid particles. However for fluid particles volumes are conserved. Thus the smallest dimension decreases so that its product with the growing area, giving the volume, stays constant. This gives $|\lambda_3|=\lambda_1+\lambda_2$ where $\lambda_1+\lambda_2$ is the growth exponent of areas defined in Eq.~(\ref{area}). The fluctuation of concentration grows until the time $t_*=|\lambda_3|^{-1}\ln(\eta/r)$ when the smallest dimension becomes equal to $r$. Beyond this time there is no growth of correlated fluctuations of concentration \cite{fphl}. We find that the factor of increase of concentration is $\exp[|\sum\lambda_i|t_*]=(\eta/r)^{D_{KY}}$ where, 
\begin{eqnarray}&& 
D_{KY}(y)=\frac{|\langle \sum \lambda_i(y)\rangle|}{|\langle \lambda_3(y)\rangle|}.
\label{eq:dky}
\end{eqnarray}
We observe that $D_{KY}$ has the structure of the reduced formula for the Kaplan-Yorke fractal codimension in the case of weak compressibility \cite{ky,fouxon1}. We defined
\begin{eqnarray}&& 
\langle \lambda_3(y)\rangle=\frac{1}{t}\int\frac{dxdz}{S} \ln\left(\frac{S(t, x, y, z)}{S(0, x, y, z)}\right),\ \  \tau_{\eta}\ll t\ll t_{in}(y), 
\end{eqnarray}
where $S(t, x, y, z)$ is an arbitrarily oriented infinitesimal area element located near $(x, y, z)$ initially. This average is independent of $t$ in the considered time interval. Further, the orientation of the surface, that can be defined by the normal, is irrelevant for the long-time limit despite the anisotropy of the statistics of turbulence. This is because orientation reaches (anisotropic) steady state quite fast \cite{review} (this is not the case for sedimenting inertial particles where relaxation of orientation is long \cite{fphl}). There is a significant difference of convergence time for $\sum\lambda_i$ and $\lambda_3$ as remarked previously: we have $\lambda_3(t, y)\approx \langle \lambda_3(y)\rangle$ at $t\gg\tau_{\eta}(y)$, cf. spatially uniform case \cite{BFF,review}.  

In the case of spatially uniform turbulence $D_{KY}$ describes all the fractal codimensions \cite{fouxon1}. For instance the pair-correlation function of concentration scales as $(\eta/r)^{\Delta}$ where $\Delta=2D_{KY}$ is the correlation codimension \cite{BFF,fouxon1}. However in the case of inhomogeneous turbulence this is no longer true. This can be seen in the simplest context by considering the exponential growth of moments of concentration which in spatially uniform case is determined by $\sum\lambda_i$ completely. We have,
\begin{eqnarray}&& 
\left\langle \left[\frac{n(t, \bm q(t, \bm x))}{n(0, \bm x)}\right]^k\right\rangle=\left\langle \exp\left[-k\int_0^t w(t', \bm q(t', \bm x))dt'\right]\right\rangle.
\end{eqnarray}
We can use the formula $\langle \exp[X]\rangle=\exp[\langle X\rangle+\langle X^2\rangle_c/2]$ for averaging of Gaussian random variable $X$ doing averaging of the last term. This can be proved using cumulant expansion and smallness of compressibility \cite{fouxon1}. We find at $t\gg \tau_{\eta}(y)$, 
\begin{eqnarray}&& 
\left\langle \left[\frac{n(t, \bm q(t, \bm x))}{n(0, \bm x)}\right]^k\right\rangle=\exp\left[-kt\left\langle\sum\lambda_i\right\rangle+k^2t\int_0^{\infty} \langle w(0)w(t)\rangle_c dt\right].
\end{eqnarray}
In the spatially uniform case described in Section \ref{sec:theory} $|\sum\lambda_i|=\int_0^{\infty} \langle w(0)w(t)\rangle_c dt$ so the growth exponents are $k(k+1)|\langle\sum\lambda_i\rangle|$. The exponent is zero at $k=-1$ because of the conservation of the number of particles \cite{BFF}. The exponents are determined by $\sum\lambda_i$ completely. In contrast, for inhomogeneous turbulence factors near $k$ and $k^2$ in the formula above become independent. The average concentration can increase or decrease locally without contradicting the global conservation of the number of particles as in turbophoresis. The growth exponents of the moments of concentration are no longer determined uniquely by $\sum\lambda_i$. The formulas derived above hold for concentration in the particle's frame. Similar formulas can be written for the growth of moments of concentration at a fixed spatial point \cite{fouxonnlin}.  

We consider statistics of particle distribution in space after transients. If we seed particles in the channel then, after transients that at scale $r$ have typical time-scale $t_*$, they distribute over a multifractal structure in space \cite{BFF,FFS,fouxon1}. The statistics of the distribution can be obtained by averaging over the $x-z$ plane as in the previous Sections. It was demonstrated in \cite{Schmidt} that pair-correlation function of concentration of particles $n(t, \bm x)$ factorizes in product of (possibly time-dependent) average local concentrations and geometrical factor of fractal increase of probability of two particles to be close. The obvious change of the formula for $w$ in \cite{Schmidt} gives, 
\begin{eqnarray}&&
\langle n(t, \bm x)n(t, \bm x+\bm r)\rangle=\langle n(t, \bm x)\rangle \langle n(t, \bm x+\bm r)\rangle\left(\frac{\eta}{r}\right)^{\Delta(y)},\ \ \ \  \frac{1}{|\lambda_3(y)|}\ln\left(\frac{\eta}{r}\right)\ll t_{in},\\&&
\Delta(y)=\frac{\tau^2}{|\lambda_3(y)|}\int_{-\infty}^{\infty}  \langle \nabla^2 p(0, y)\nabla^2 p(t)\rangle_c dt=2D_{KY}(y)+\frac{\partial_y v_{eff}(y)}{|\lambda_3(y)|}.
\label{eq:delta}
\end{eqnarray}
We see that the correlation codimension is not $2D_{KY}$. It scales proportionally with $St^2$ when $D_{KY}$ has both the term that scales linearly and the terms that scales quadratically with $St$. The reason why the $\partial_y v_{eff}(y)$ term in $\sum\lambda_i$ drops from $\Delta$ is that this term originates in the average velocity that affects equally the average $n$ and its fluctuations disappearing from the ratio $n/\langle n\rangle$. The form of $\langle n(t, \bm x)\rangle$ in Eq.~(\ref{eq:delta}) is determined by initial and boundary conditions on the concentration and is problem-dependent in contrast with the power-law factor. Further the correlations do not depend on the direction of $\bm r$ despite anisotropy of the statistics of turbulence. This is restoration of isotropy that originates in independence of Lyapunov exponents on the initial orientations. 

Finally we consider the coarse-grained concentration $n_l(\bm x)$,
\begin{eqnarray}&&
n_l(\bm x)=\frac{m_l(\bm x)}{4\pi l^3/3},\ \ m_l(\bm x)=\int_{|\bm x'-\bm x|<l}n(0, \bm x') d\bm x',
\end{eqnarray}
which is mass $m_l(\bm x)$ in small volume of radius $l\ll \eta$ divided by the volume. We can find $m_l(\bm x)$ using the consideration of \cite{fouxon1,Schmidt} by tracking the ball of the particles back in time to time $t=-t_*$ where $t_*=-|\lambda_3(y)|^{-1}\ln(\eta/l)$. Since there are no fluctuations of concentration at scale $\eta$ then the mass of the ball at that time is volume $(4\pi l^3/3)\exp[-\int_{-t_*}^0 w(t', \bm q(t', \bm x))dt']$ times the average concentration $\langle n(-t_*, \bm q(-t_*, \bm x))\rangle$ see details in \cite{fouxon1}. Comparing the resulting formula for $n_l(\bm x)$ with the formula for $\langle n(\bm x, t)\rangle$ from \cite{Schmidt} we find 
\begin{eqnarray}&&
\frac{n_l(\bm x)}{\langle n(\bm x, t)\rangle}=\frac{\exp[-\int_{-t_*}^0 w(t', \bm q(t', \bm x))dt']}{\left\langle\exp[-\int_{-t_*}^0 w(t', \bm q(t', \bm x))dt']\right\rangle}.
\end{eqnarray}
This formula was provided (with a typo) in \cite{Schmidt} where the result of the averaging was presented, 
\begin{eqnarray}&&
\frac{\langle n_l^k(\bm x)\rangle}{\langle n(\bm x, t)\rangle^k}=\left(\frac{\eta}{l}\right)^{\Delta(y)k(k-1)/2},
\end{eqnarray}
which for $k=2$ corresponds with the previously derived formula for the pair-correlation of concentration. Thus if we use the scaling of $n_l/\langle n(\bm x, t)\rangle$ for defining fractal dimensions \cite{fouxon1} then none of the dimensions is $D_{KY}$.

\section{Numerical Simulation}\label{sec:dns}

We use the direct numerical simulation (DNS) of a turbulent channel flow provided by the JHTDB. A large variety of time and space dependent, Eulerian simulation results are stored on a cluster of databases, which is made accessible to the public. The functionality of this database systems and details on the simulations available, as well as confirmations of their validity, are described in \cite{perlman2007data,li2008public,yu2012studying,graham2016web} and other references therein. All details on the DNS computation, specifically numerical schemes, discretization methods and further simulation details of the turbulent channel flow are extensively described in \cite{graham2016web}.


The turbulent channel flow with a friction Reynolds number $Re_{\tau}\approx1000$, considered in this work is a wall bounded flow with no-slip conditions at the top and bottom walls ($y/h=\pm1$, where $h$ corresponds to half of the channel height) and periodic boundary conditions in the longitudinal and transverse directions. In this channel flow DNS, the streamwise direction $x$ and the transverse direction $z$ can be considered homogeneous, whereas the wall-normal direction $y$ serves as inhomogeneous direction for the purpose of generating turbophoretic drift of inertial particles. The domain spans over the three directions as follows: $L_x \times L_y \times L_z = 8\pi h\times 2h\times 3\pi h$, where $h=1$ in dimensionless units. Quantities normalized by the friction velocity $u_{\tau}$, the viscous length $\nu/u_{\tau}$ ($\nu=$ viscosity) or the viscous time scale $\nu/u_{\tau}^2$ are presented with the superscript $+$. The wall of the channel is located at $y^+=0$, the center of the channel is at $y^+=1000$. 
An overview over the main simulation, flow and grid parameters is given in Table \ref{table:parametes}.

\begin{table}
\begin{tabular}{K{2.6cm} K{2.8cm} K{1.2cm} K{1.2cm} K{1cm} K{1cm} K{1.1cm} K{1cm} K{1.1cm} K{1.1cm} K{1cm} K{1cm}}
\hline\hline \noalign{\vskip 2mm}
$L_x\times L_y\times L_z$ &$N_x\times N_y\times N_z$ &  $\delta t$ & $\nu$ & $U_c$ &  $u_{\tau}$ & $Re_{\tau}$ &$St^+$ & $\Delta x^+$ & $\Delta y_1^+$ & $\Delta y_c^+$ & $\Delta z^+$ \\ \hline \noalign{\vskip 2mm}	
$8h\pi\times 2h\times 3h\pi $ & $2048\times 512\times 1536$ & $0.0065$ & $5\times10^{-5}$ & $1.13$ & $0.0499$ & $999.4$ & $1$ & $12.26 $ & $0.0165$ & $6.16$ & $6.13$\\
   [0.25cm] \hline\hline \noalign{\vskip 2mm}																							
\end{tabular}
\caption{\label{table:parametes} Simulation, flow and grid parameters - $L_{i}$: domain size in all directions $i=x,y,z$, $N_{i}$: number of grid points in all directions, $\delta t$: time step, $\nu$: viscosity, $U_c$: centerline velocity, $u_{\tau}$: friction velocity, $Re_{\tau}$: friction Reynolds number, $St^+$: viscous Stokes number, $\Delta x^+$: grid spacing streamwise direction, $\Delta y_1^+$: grid spacing wall-normal direction (first point), $\Delta y_c^+$: grid spacing wall-normal direction (central point), $\Delta z^+$: grid spacing spanwise direction}
\end{table}

The time step, $\delta t$, at which the Eulerian flow data can be extracted from the database is $0.0065$ and the total available flow time is approximately $26$ (non-dimensional time units), which corresponds to approximately one flow through time. The total duration of the simulation is thus $t_{end}^+=2.6\times10^4$.

We have used the Eulerian results of the channel flow DNS provided by the JHTDB to perform Lagrangian tracking of inertial particles in the channel flow. At total number of, $4\times10^6$ point-particles are randomly seeded across the entire channel domain. The JHTDB allows to extract velocity, velocity gradients and the Hessian of pressure at any arbitrary particle position. We use Eq.\,(\ref{eq:particlevelocity}) to determine the inertial particle velocity. We compute the second term on the RHS of Eq.\,(\ref{eq:particlevelocity}) $(\partial_t\bm u+(\bm u\cdot\nabla)\bm u)$ based on the material derivative $D\bm u/Dt$ along a tracer particle trajectory. This is done by applying a simple finite difference scheme using the tracer particle velocity at the tracer particle position of two consecutive time steps. The inertial particles are advected with time step $\delta t$ applying a second order Adams-Bashforth method for temporal integration. 

\begin{figure}[h]
\centering
\includegraphics[scale=0.37]{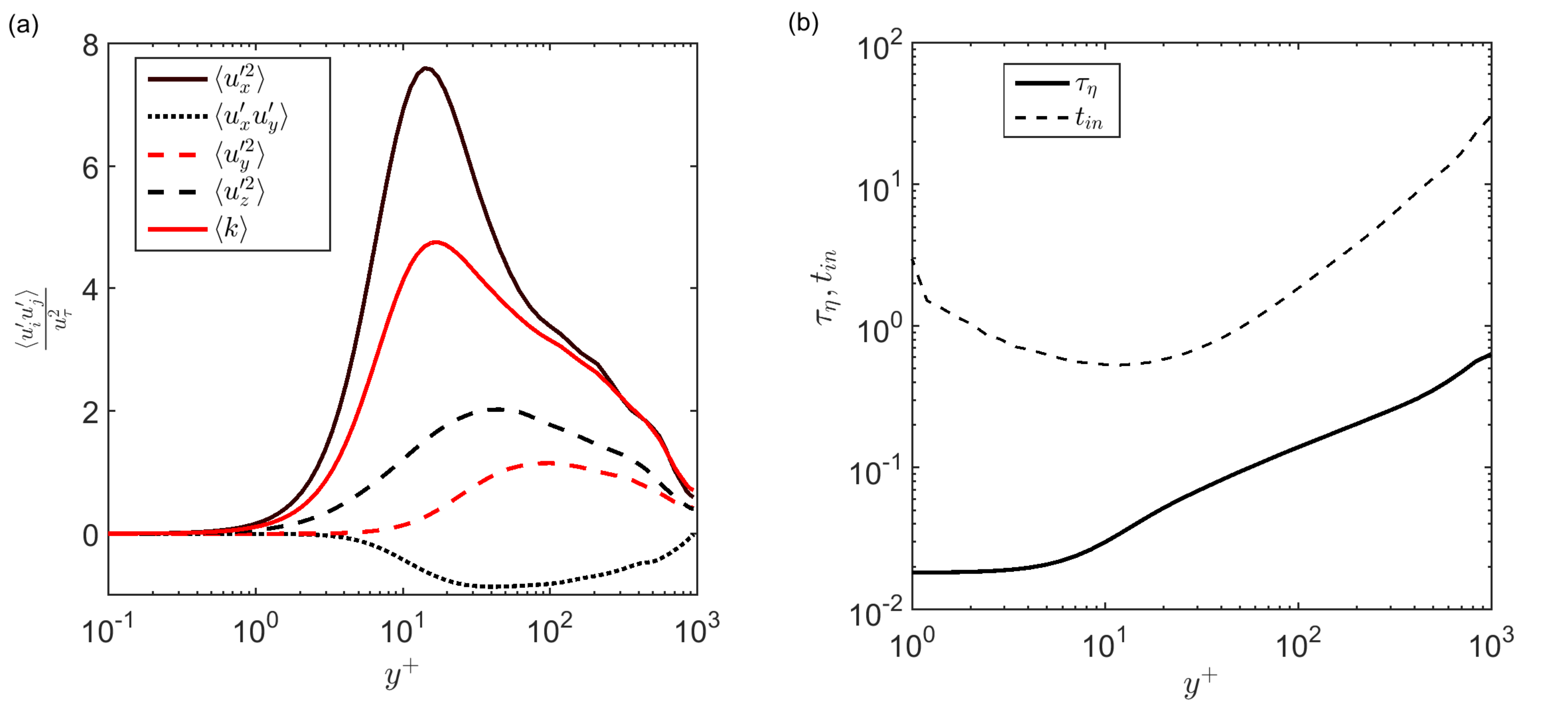}
\caption{\label{fig:Ch_restress_full}(a) - Components of the Reynolds stress tensor and turbulent kinetic energy $k$ normalized by $u_{\tau}^2$ plotted along non-dimensional channel height. The turbulent kinetic energy (red solid line) peaks at $y^+=10-11$. The square of the wall-normal velocity fluctuations $\langle u'_yu'_y\rangle/u_{\tau}^2$ (red dashed line) has its maximum at $y^+=100$. (b) - Variation of the the Kolmogorov time-scale $\tau_{\eta}$ (solid line) and the inhomogeneity time $t_{in}=y/\langle u_y^2\rangle^{1/2}$ (dashed line) along the channel height.} 
\end{figure} 

A verification of the channel flow DNS, including a comparison with previous works, has been done in \cite{graham2016web}. We show the Reynolds stresses $u'_iu'_j$ as well as the turbulent kinetic energy, normalized by the friction velocity $u_{\tau}$ versus the viscous wall distance $y^+$ in Fig.\,\ref{fig:Ch_restress_full}(a) and plot the wall-distance logarithmically in order to focus the visualization on the near-wall behavior of the flow. The dashed red line, indicating the Reynolds stresses based on the wall-normal velocity $u'_y$, shows a maximum at around $y^+=100$ dropping quite steeply towards the wall. We highlight this term in Fig.\,\ref{fig:Ch_restress_full}, since its second derivative  is responsible for the turbophoretic migration of the particles as shown previously in Eq.\,(\ref{eq:divv_inh}).

The ratio between the particle response time $\tau$ and the Kolmogorov time scale $\tau_{\eta}$, defines the particle Stokes number $St$. Due to the strong dependence of the turbulent kinetic energy on the wall-normal direction (red solid line in Fig.\,\ref{fig:Ch_restress_full}(a)), also the dissipation of the turbulent kinetic energy and thus the Kolmogorov microscales, depend on $y$. Fig.\,\ref{fig:Ch_restress_full}(b) shows $\tau_{\eta}$ versus the inhomogeneous spatial direction $y^+$. For the presented theory to be valid we choose a rather weak particle inertia by setting the Stokes number averaged over the whole channel to $\langle St\rangle=0.1$, based on the averaged Kolmogorov time $\langle\tau_{\eta}\rangle=0.2133$. This yields $\tau=0.0213$ and thus a viscous Stokes number of $St^+\approx1$. The viscous Stokes number is defined in terms of the friction velocity $u_{\tau}$ as $St^+=\tau u_{\tau}^2/\nu$. Due to the variation of $\tau_{\eta}$ along wall-normal direction (Fig.\,\ref{fig:Ch_restress_full}) also the local $St$ depends strongly on $y$, reaching the largest values in the vicinity of the wall.

\section{Results and Discussion}\label{sec:analysis}
\subsection{Results}
\begin{figure}[h]
\centering
\includegraphics[scale=0.5]{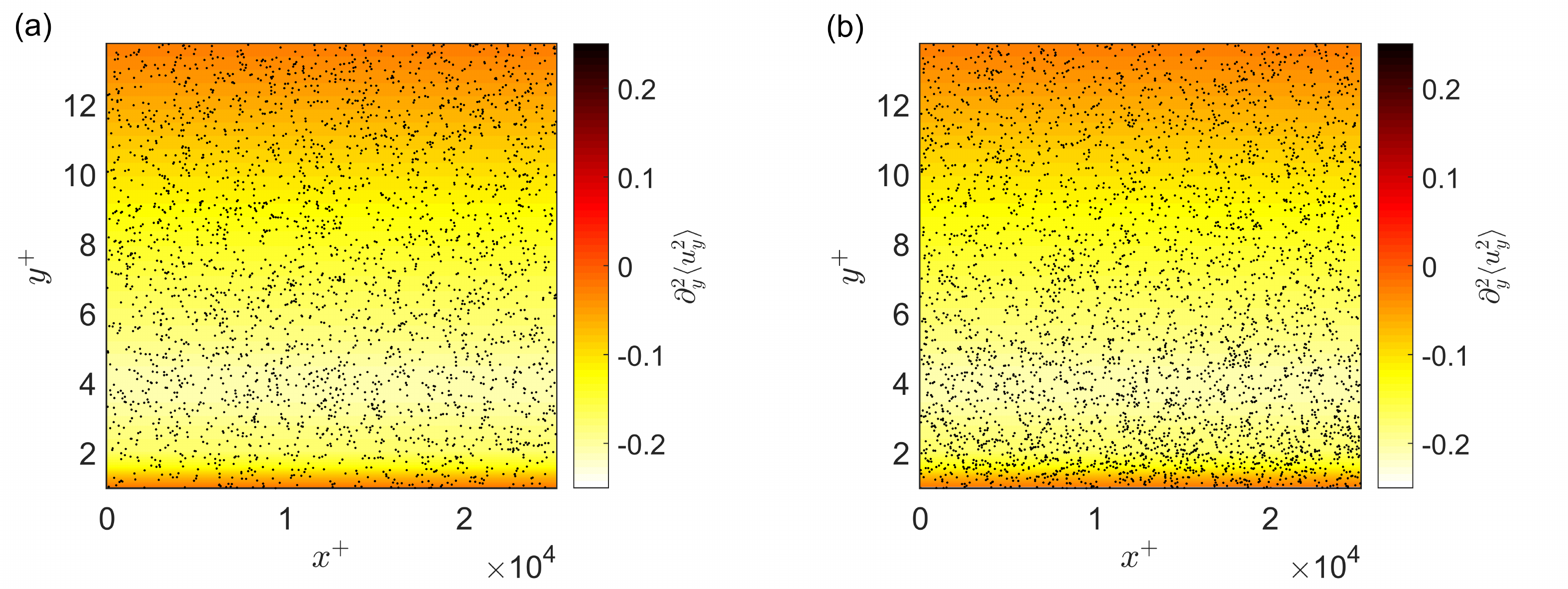}
\caption{\label{fig:particledistribution} Comparison of the instantaneous particle distribution of tracer particles ($St^+=0$ - plot a) and inertial particles ($St^+=1$ - plot b) at time $t^+=2.5\times10^4$ in the near-wall region between $y+=0-15$ (dimensions are not to scale). The particles are plotted on top of the time-averaged $\partial_y^2\langle u_y^2\rangle$ field.}
\end{figure}

Before going into the analysis of cluster formation, we start with discussing the effect of turbophoresis. In a channel flow inertial particle migration due to turbophoresis changes the initially random distribution of particles by driving particles towards the wall. A snapshot of the distribution of tracer and inertial particles at $t^+=2.5\times 10^4$ in the $x-y$ plane near the wall (Fig.\,\ref{fig:particledistribution}) shows the increased inertial particle concentration near the wall qualitatively. The plane chosen here is located at $z=1.5\pi$ and covers the full extension of the channel in streamwise direction but only the near wall region between $y^+=0$ and $y^+=15$. Particles are plotted as black points on top of the time-averaged $\partial_y^2\langle u_y^2\rangle$ field. It is  visible that the tracer particles shown in Fig.\,\ref{fig:particledistribution}(a) distribute randomly in space and appear uniformly distributed while inertial particles (Fig.\,\ref{fig:particledistribution}(b)) accumulate near the wall. 

\begin{figure}[h]
\centering
\includegraphics[scale=0.37]{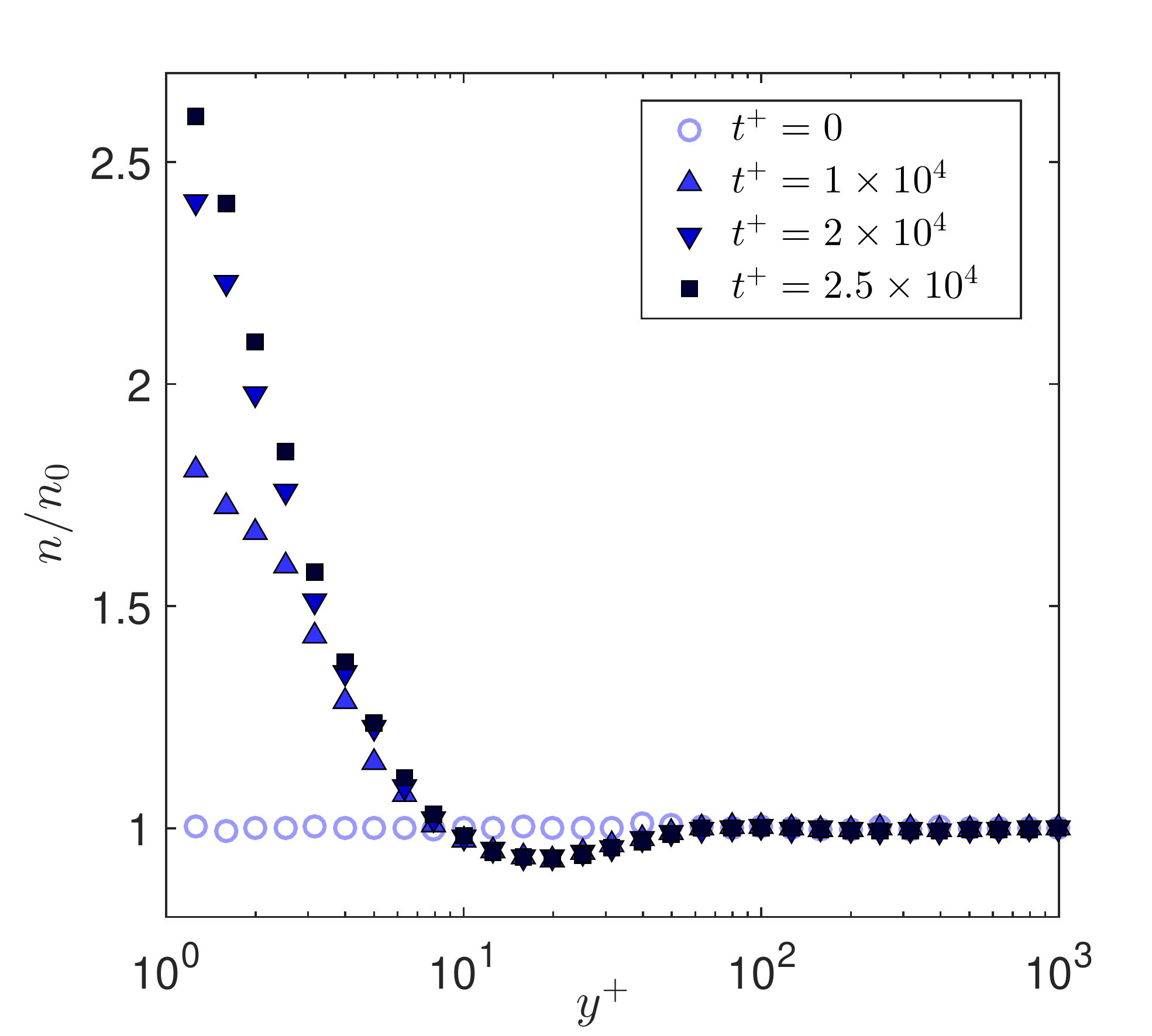}
\caption{\label{fig:particleconcentration} Particle concentration $n$ normalized by the initial, random, particle concentration $n_0$ versus the wall-normal distance $y^+$. Different symbols indicate the temporal evolution from $t^+=0$ to $t^+=2.3\times10^4$. The region below $y^+=10$ shows an increase in particle concentration, whereas the particle concentration drops below the initial concentration between $y^+=10$ and $y^+=60$.}
\end{figure}

The temporal evolution of wall-normal particle concentration profiles are depicted in Fig.\,\ref{fig:particleconcentration}, which compares  the particle concentration $n$  at the initial time step $t^+=0$ (circles), as well as $t^+=10^4$ (triangle), $t^+=2\times10^4$ (triangle upside down) and $t^+=2.5\times10^4$ (squares), normalized by the initial, random particle concentration $n_0$. It is visible that with time more particles accumulate in the vicinity of the wall. In the region below $y^+=10$ a significant increase of the particle concentration is observable. In the vicinity of the wall, the ratio $n/n_0$ rises almost up to factor three in the considered time span, despite the relatively weak inertia of the particles. Above $y^+=10$ the particle concentration decreases below the initial concentration where a minimum of $n/n_0=0.95$ is reached at $y^+=20$. This leads to a specific interest of the degree of clustering in both, the region where the number of particles is high (below $y^+=10$) and the region where the number of particles is low (between $y^+=10-70$). 

The degree of clustering is not only determined by the particle concentration itself but instead rather by a high probability of finding particles with very small inter-particle distance. As explained in section \ref{sec:theory}, particles approach each other and thus form clusters if $\nabla\cdot\bm v<0$ along particle trajectories. This effect is quantified by the previously introduced sum of Lyapunov exponents, $\sum\lambda_i$.
Negative values of $\sum\lambda_i$ correspond to a compression of infinitesimal volumes formed by particles, whereas positive values of $\sum\lambda_i$ correspond to diverging volumes. For a precise quantification of the degree of clustering arising from the combined action of inhomogeneous and homogeneous clustering effects, it is necessary to compute the finite-time Lyapunov exponents. We do this in the following by the computation of each individual term of the RHS in Eq.\,(\ref{sum}) or Eq.\,(\ref{complete}), respectively.




Since all these terms depend on time, it is important to look at the convergence time of the individual terms. Theoretically, convergence within a few Kolmogorov time scales of the first term ($\tau\partial_y\int_0^t\langle u_y(0)\nabla^2p(t')\rangle_c dt'$) as well as the second term ($-\tau^2\int_0^t\langle\nabla^2p(0)\nabla^2p(t')\rangle_c dt'$) on the RHS of Eq.(\ref{complete}) is expected. These two integrals use the cumulant terms as described in Eq.\,(\ref{eq:cumulants}). As discussed in section \ref{sec:lyapunov} the third integral, which is $\int_0^t\langle\nabla^2p\rangle$, does not converge. However, at the time where the other terms have converged the full term $\tau^2(\partial_y\langle u_y^2\rangle)\partial_y\int_0^t\langle\nabla^2p(t')\rangle dt'$ (including this integral) remains small and can therefore be excluded from the computation of $\sum\lambda_i$ as shown below. For the approximations to be valid the inhomogeneity time $t_{in}$ (Fig.\,\ref{fig:Ch_restress_full})(b) has to be larger than the convergence time of these integrals.

\begin{figure}[h]
\centering
\includegraphics[scale=0.45]{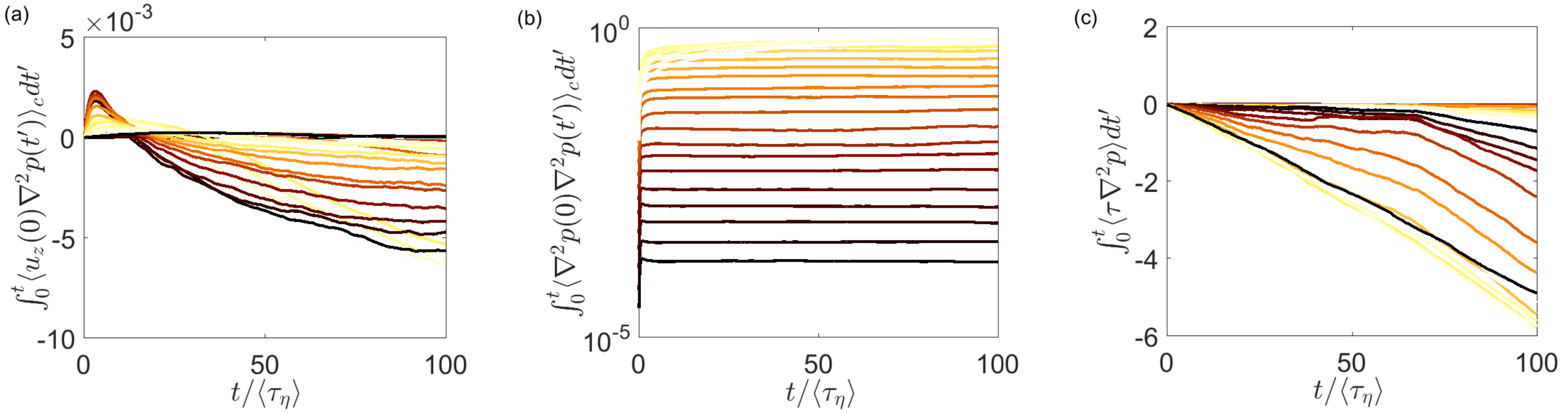}
\caption{\label{fig:integrals} The three integrals of Eq.\,(\ref{sum}) are plotted versus time normalized by the average Kolmogorov time for all wall distances used for the computation. Darker curves present the regions in the center of the channel and lighter curves the near-wall region. (a): $\tau\partial_y\int_0^t\langle u_y(0)\nabla^2p(t')\rangle_c dt'$  (b): $-\tau^2\int_0^t\langle\nabla^2p(0)\nabla^2p(t')\rangle_c dt'$ (c): $\int_0^t\langle\nabla^2p(t')\rangle dt'$ }
\end{figure}

The temporal evolution of the three integrals of Eq.\,(\ref{sum}) in Fig.\,\ref{fig:integrals}. The color shading is darker for increasing $y^+$, i.e. light curves refer to regions near the wall and dark curves in the bulk of the channel, respectively. The integral of the first term on the RHS of Eq.\,(\ref{sum}) reaches relatively low values in the viscous sublayer and in the bulk, whereas it becomes more significant in the intermediate (log-layer) regions where convergence takes longer. In the bulk as well as in the viscous sublayer the curves converge to relatively low values.  
The second integral of Eq.\,(\ref{sum}) is shown in Fig.\,\ref{fig:integrals}(b), where it is seen that all curves converge fast. The largest values are found in the regions between $y^+=6-25$, where turbulence intensity peaks (see Fig.\,\ref{fig:Ch_restress_full}(a)). Figure \,\ref{fig:integrals}(c) shows $\int_0^t\langle\nabla^2p(t')\rangle dt'$ that as expected does not converge. Generally, the convergence time is smaller than $t_{in}$. However, as one can see from Fig.\,\ref{fig:integrals}(a) in the buffer layer region the convergence of the first term is rather slow and an upper limit can be estimated at about $100\langle\tau_{\eta}\rangle$, which is the time we choose to evaluate the mean Lyapunov exponents.

\begin{figure}[h]
\centering
\includegraphics[scale=0.49]{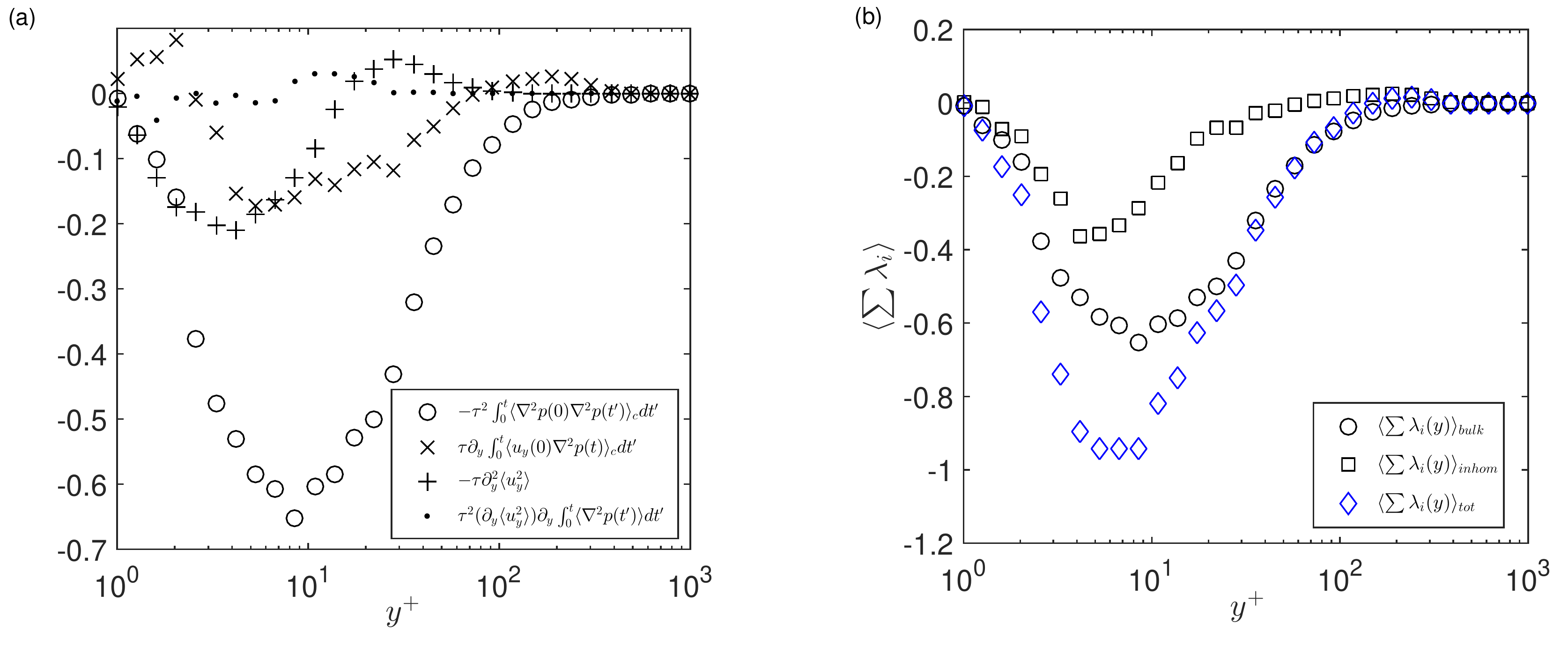}
\caption{\label{fig:3termcorrelation}:(a) All four terms of the RHS of Eq.\,(\ref{complete}) versus channel height; $-\tau^2\int_0^T\langle \nabla^2p(0)\nabla^2p(t)\rangle dt$ - circles; $-\tau\nabla_z^2\langle u_z^2\rangle$ - crosses; $\tau\nabla_z \int_0^T\langle u_z(0)\nabla^2p(t)\rangle dt$ - stars; $\tau^2(\partial_y\langle u_y^2\rangle)\partial_y\int_0^t\langle\nabla^2p(t')\rangle dt'$ - black points.   (b): We show $\langle\sum\lambda_i\rangle$ separately for the homogeneous (circles) and the inhomogeneous contribution (squares), consisting of the sum of $-\tau\nabla_z^2\langle u_z^2\rangle$ and $\tau\nabla_z \int_0^T\langle u_z(0)\nabla^2p(t)\rangle dt$ but not including the last term of Eq.\,(\ref{complete}) The sum of the two components $\langle\sum\lambda_i\rangle_{tot}$, according to Eq.\,(\ref{eq:29}) is presented as blue diamonds.}
\end{figure}

Now we evaluate the individual terms on the RHS of Eq.\,(\ref{complete}) in Fig.\,\ref{fig:3termcorrelation}(a). Circles show the wall-normal profile of the first term which is the sole term causing small-scale clustering of inertial particles in homogeneous turbulence.

 This correlation is zero at the wall but its absolute value increases further away from the wall reaching the largest negative values in the range between $y^+=5-20$. Beyond $y^+=20$ the curve approaches small magnitudes as turbulence becomes more homogeneous but will not vanish.  

The second term on the RHS of Eq.\,(\ref{complete}) ($\tau\nabla_y\int_0^T\langle u_y(0)\nabla^2p(t)\rangle$ - x symbols) is negative and contributes to clustering in a part of the buffer layer, whereas it is positive (and counteracting clustering) outside that region. 

The third term of Eq.\,(\ref{complete}) ($-\tau\nabla_y\langle u_y^2\rangle$), shown as $+$ symbols, accounting for the turbophoretic effect solely is negative below $y^+=15$. This indicates a compression of infinitesimal volumes and contribution towards clustering (this location has been defined as $y*$ in section \ref{sec:theory}). The curve changes sign contributing against clustering above $y^+=15$. Above $y^+=15$ the term becomes slightly positive before converging towards $0$ at $y^+=100$. 

The fourth term of Eq.\,(\ref{complete}) shown as filled black points in Fig.\,\ref{fig:integrals}(a) is rather small at the time where the other terms have reached convergence as predicted and will be neglected in the remaining analysis below.

In Fig.\,\ref{fig:3termcorrelation}(b) the terms of Eq.\,(\ref{eq:29}) are displayed. We divide the total $\sum\lambda_i$ (diamond symbols) in a homogeneous or bulk component (circles) and an inhomogeneous component (squares). Despite the linear dependence of the inhomogeneous component on $St$ the clustering due to the homogenous contribution generally exceeds the inhomogeneous component. Both components add up and reach the maximal negative $\sum\lambda_i\approx -1$ at $y^+=6$.

\begin{figure}[h]
\centering
\includegraphics[scale=0.49]{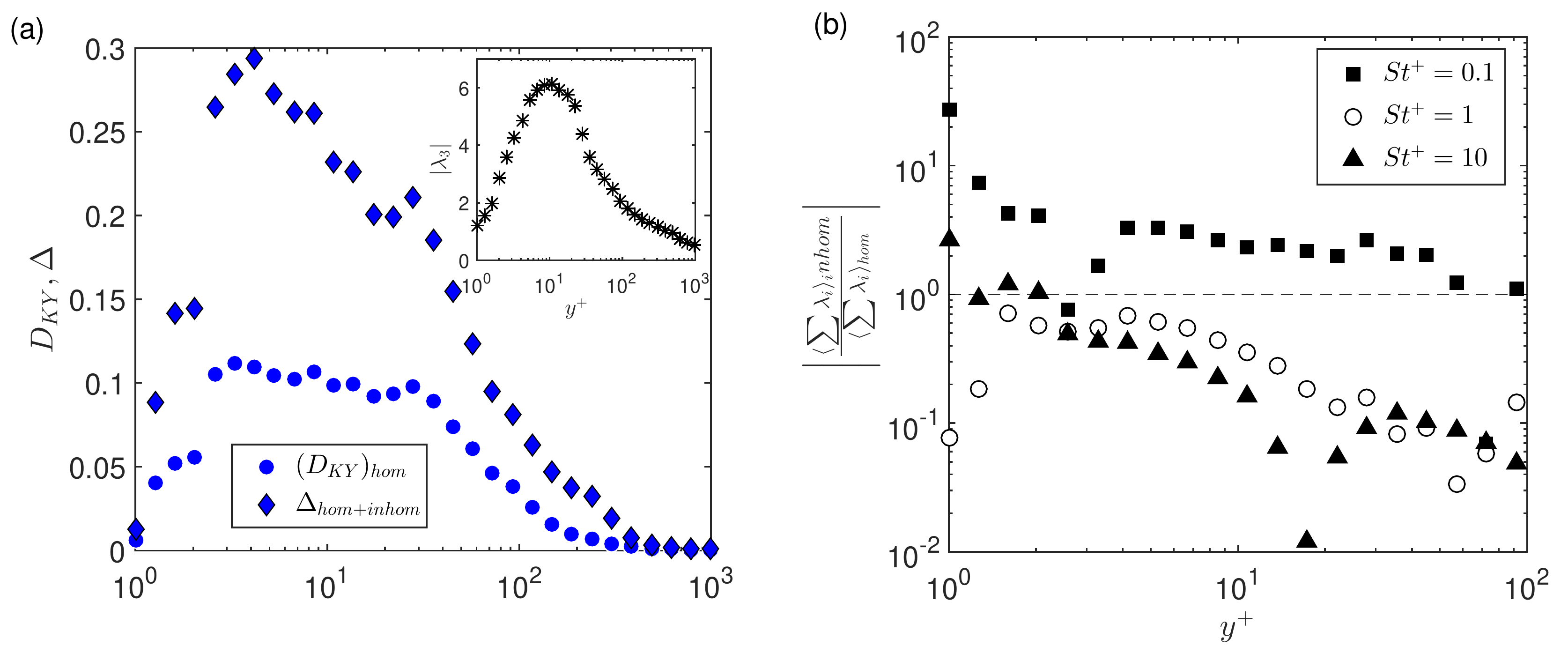}
\caption{\label{fig:lambda_3} The blue circles show $(D_{KY})_{hom}$, obtained from Eq.\,(\ref{eq:dky}) over the wall-normal distance showing a strong increase in the near-wall region and the largest values in the region between $y^+=2-8$. The diamonds indicate the correlation co-dimension $\Delta_{hom+inhom}$, computed according to Eq.\,(\ref{eq:delta}), varying over wall-distance. Inset: Dependence of $\lambda_3$, computed according to the procedure described above, on the viscous wall distance $y^+$. The largest value can be observed at $y^+=10$ (b): Plot of $\left|\frac{\langle\sum\lambda_i\rangle_inhom}{\langle\sum\lambda_i\rangle_{hom}}\right|$ form $y^+=1$ to $y^+=100$ for three different Stokes numbers - $St^+=0.1$ (filled squares); $St^+=1$ (empty circles); $St^+=10$ (filled triangles). Note that the strong outliers in a few positions result from a change of sign of the inhomogeneous term.}
\end{figure}

Since we aim to quantify the preferential concentration via $D_{KY}$ and the correlation codimension $\Delta$ according to Eq.\,(\ref{eq:dky}) and (\ref{eq:delta}) respectively, in dependence on the inhomogeneous flow direction, the third Lyapunov exponent $\lambda_3$ has to be computed. The calculation of $\lambda_3$ is performed via the finite-time Lyapunov exponents that provide a measure of the cumulative deformation of the particles \cite{meneveau2016analysis}. This requires an estimation of the instantaneous deformation rate of the particle trajectory, which is done via the instantaneous Lyapunov exponents $\lambda'_{ii}$. These instantaneous Lyapunov exponents can be computed by the alignment of the eigenvectors of the Cauchy-Green tensor $C_{ij}$ of a particle trajectory and the velocity gradient tensor. More precisely the instantaneous Lyapunov exponents at each particle location can be found, using

\begin{equation}
\lambda'_{ii}=cos^2(\theta_{ij})P_j ,
\label{eq:FLTLE}
\end{equation}
where $P_j$ are the eigenvalues of the strain rate tensor and $\theta_{ij}$ is the angle between eigenvector $i$ of the Cauchy-Green tensor and eigenvector $j$ of the strain rate tensor. Averaging those instantaneous Lyapunov exponents along a Lagrangian path enables us to determine $\lambda_3$ depending on the wall distance $y$ of the channel. The corresponding result is presented in the inset of Fig.\,\ref{fig:lambda_3}(a). The maximum $|\lambda_3|$ is reached at $y^+=10$, the curve drops quite fast in both directions. As apposed to spatially uniform turbulence the estimate that $\lambda_3\approx\tau_{\eta}^{-1}$ is not true along the entire channel. Below $y^+=100$ the product of $\tau_{\eta}\lambda_3$ drops significantly, whereas it is constant above $y^+=100$.

The clustering introduced by $(\langle\sum\lambda_i\rangle)_{bulk}$ is quantified using $(D_{KY})_{hom}$. The resulting values are shown as blue circles in Fig.\,\ref{fig:lambda_3}(a). With $(D_{KY})_{hom}=0.1$ the maximum level is moderately high but interestingly it stays almost constant across a large region ranging from $y^+=3$ to $y^+=30$. 
The values obtained for $\Delta_{hom+inhom}$, combining homogeneous and inhomgeneous clustering, are relatively high in regions below $y^+=10$, see Fig.\,\ref{fig:lambda_3}(a). At the wall $\Delta$ is $0$ but starts increasing rapidly. The curve peaks around $y^+=4$ with $\Delta_{hom+inhom}\approx 0.3$. It then decreases slowly to almost $0$ at $y^+=200$. In contrast to $(D_{KY})_{hom}$ the correlation codimension takes on very large values in a much narrower range from $y^+=2-10$.  

One key difference between the clustering in homogeneous and inhomogeneous turbulence is that for weakly inertial particles there is a linear dependence on St (for the first term on the RHS of Eq.\,(\ref{complete})). The ratio $\left|\frac{\langle\sum\lambda_i\rangle_inhom}{\langle\sum\lambda_i\rangle_{hom}}\right|$ changes not only significantly throughout the channel height but also for different Stokes numbers. One would expect this term to dominate in inhomogeneous regions of the flow. In the case examined in this study so far, this is not the case. The homogeneous term is generally higher or at least of the same magnitude as the inhomogeneous term. Therefore, we want to extend this study to other $St$. In Fig.\,\ref{fig:lambda_3}(a) we show the ratio between the homogeneous and the inhomogeneous contributions to clustering  additionally for the case where $\langle St\rangle=1$ or $St^+=1$ (empty circles) and $St^+=0.1$ (filled squares) and $St^+=10$ (filled triangles).  
We find that the Stokes number has a strong impact on which term dominates the clustering. However, predicting the behavior is not trivial due to the variations of the different terms along the $y-$ direction. As mentioned before, for $St^+=1$ the ratio of $\left|\frac{\langle\sum\lambda_i\rangle_inhom}{\langle\sum\lambda_i\rangle_{hom}}\right|$ stays below $1$ throughout the entire channel. Interestingly, for particles with small inertia the inhomogeneous term will dominate, as can be seen for the case of $St^+=0.1$ in Fig.\,\ref{fig:lambda_3}(b). For particles with larger inertia the behavior becomes more complex. The inhomogeneous terms will dominate or be of the same order as the homogeneous terms right at the wall but become less important throughout the rest of the channel. The homogeneous terms dominate even more strongly than in the case of $St^+=1$ further away from the wall.

\subsection{Discussion}
The results presented in the previous section indicate a strong dependence of particle concentration and clustering on the wall-normal direction. Turbophoresis in a turbulent channel flow drives particles towards the wall. After $t^+=2.5\times10^4$ the initial concentration at the wall is exceeded by almost factor 3 within the considered time span. The turbophoretic particle migration is driven by $-\tau\nabla_z^2\langle u_z^2\rangle$ shown in Fig.\,\ref{fig:3termcorrelation}(a). The change of sign in this term at $y^+\approx 15$ determines the wall-distance below which particles start accumulating and above which the particle concentration decreases. 

We analyze the space-dependent rate of creation of inhomogeneous particle concentration to quantify inertial particle clustering in inhomogeneous turbulence. All three terms investigated according to Eq.\,(\ref{sum}) and Eq.\,(\ref{complete}) determining $\langle\sum\lambda_i\rangle$, depend differently on the wall distance. The term $\tau^2(\partial_y\langle u_y^2\rangle)\partial_y\int_0^{\infty}\langle\nabla^2p(t')\rangle dt'$ was small enough to be neglected for times within the convergence time of the other terms. Particle clustering for the case of $St^+=1$ is dominated by the homogeneous fractal clustering, represented by $-\tau^2\int_0^t\langle\nabla^2p(0)\nabla^2p(t')\rangle_c dt'$ even close to the wall where inhomogeneity is strongest. Its dependence on $y$ (Fig.\,\ref{fig:3termcorrelation}) is similar to the one of the turbulent kinetic energy, shown in Fig.\,\ref{fig:Ch_restress_full}(a). In regions of strong turbulence, $\tau\nabla^2p$ takes on large values and causes stronger clustering. Despite the linear dependence on $\tau$ the inhomogeneous contribution $(\langle\sum\lambda_i\rangle)_{inhom}$ to the overall clustering degree is smaller than $(\langle\sum\lambda_i\rangle)_{bulk}$ in the case of $St^+1$. Though $(\langle\sum\lambda_i\rangle)_{inhom}$ behaves similarly to $(\langle\sum\lambda_i\rangle)_{bulk}$ its absolute value peaks already at $y^+=4-5$ and is below the peak of $(\langle\sum\lambda_i\rangle)_{bulk}$. As can be seen in Fig.\,\ref{fig:lambda_3} the significance of the inhomogeneous terms decreases with larger wall-distance. 
We use the correlation codimension of the multifractal formed by inertial particles to quantify the strength of particle clustering. The values found for the the correlation codimension ($\Delta$) for the case under investigation here are relatively large for particles with such small inertia. In the region between $y^+=2$ to $y^+=10$, $\Delta$ reaches up to $0.3$. In the region between $y^+=10-70$ --- where the particle concentration drops below the initial concentration --- the correlation codimension decreases rapidly. Whereas, the local homogeneous turbulence contribution to clustering $(D_{KY})_{hom}$ stays almost constant at $(D_{KY})_{hom}=0.1$  from  $y^+=3-30$. Studies with particles of similar inertia find lower values of $\Delta$ (between $10^{-2}$ and $10^{-1}$) in homogeneous isotropic turbulence, e.g. \cite{saw1,Collinstwo}. This shows that turbulent inhomogeneity enhances the clustering degree particularly in the viscous sublayer and the onset of the buffer layer for $St^+=1$. The difference between $\Delta$ and $D_{KY}$ shows also that the inhomogeneous terms enhance clustering but also affect the regions where clustering occurs by the peak of $\Delta$ towards the wall. This can be explained by the maximal absolute values of $(\langle\sum\lambda_i\rangle)_{inhom}$ and $(\langle\sum\lambda_i\rangle)_{bulk}$ in Fig.\,\ref{fig:3termcorrelation}(b). Therefore, a complete investigation, unifying the clustering effects of both mechanism is essential for a precise quantification of the preferential concentration in inhomogeneous turbulent flows.

Due to the linear dependence of $(\langle\sum\lambda_i\rangle)_{inhom}$ on $\tau$ and the quadratic dependence of $(\langle\sum\lambda_i\rangle)_{bulk}$ on $\tau$, the situation changes for different particle inertia. In Fig.\,\ref{fig:lambda_3}(a) we saw that for particles with even smaller inertia $St^+=0.1$ the inhomogeneous terms dominate clustering throughout the entire channel. Instead for particles with large inertia ($St^+=10$) the inhomogeneous terms dominate near the wall but then become outweight by the homogeneous clustering term. The complementary effects of inhomogeneous and homogeneous clustering cause a different degree of clustering depending on inertia and the wall-normal direction.

\section{Conclusion}\label{sec:discussion}
In this study the combined effects of turbophoresis and small-scale fractal clustering of weakly inertial particles have been analyzed theoretically as well as numerically. A novel theoretically approach allows to describe the preferential concentration resulting from homogeneous and inhomogeneous particle clustering in inhomogeneous turbulence. We determine a space-dependent rate ($\langle\sum\lambda_i(y)\rangle$) that create inhomogeneities of concentration of particles. We find that $\sum\lambda_i$ depends linearly on the particle Stokes number, as opposed to homogeneous turbulence where $\sum\lambda_i\propto St^2$. The theoretical predictions for the creation of particle inhomogeneities have been investigated by direct numerical simulations of a turbulent channel flow using the JHTDB. The results reveal a strong turbophoretic migration of particles, despite the relatively small inertia. We find that the clustering degree depends strongly on the wall-normal direction. The strongest effects of preferential concentration can be observed in the transition of the viscous sublayer to the buffer layer at $y^+=2-10$, where local homogeneous terms contribute to clustering and clustering due to the inhomogeneity of the flow is strong. The correlation codimension rises up to $0.3$ which is remarkably high for particles with relatively weak inertia of $St^+=1$ or $\langle St\rangle=0.1$.  The values found for $\Delta$ in the near-wall region are larger compared to what particles with the same inertia would show in homogeneous turbulence. The contributions to clustering from homogeneous and inhomogeneous terms varies strongly with the wall-distance but as we show also with particle inertia. For particles with very small inertia the inhomogeneous terms outweigh the homogeneous contributions significantly. However, for particles with large inertia homogeneous turbulence will mainly determine clustering except in the vicinity of the wall, where the contribution of inhomogeneous turbulence might be stronger. 
 The findings of this work allow for a precise quantification of inhomogeneous preferential concentration of weakly inertial particles in non-uniform turbulent flows. The case of a turbulent channel flow investigated here, serves as general example for all inhomogeneous turbulent flows. Thus, the presented results can be easily transferred to investigate preferential concentration of weakly inertial particles in other frequently occurring turbulent flows, e.g. pipe or free shear flows.


\section{Acknowledgments}

L.S. would like to thank Stephen S. Hamilton for his support regarding the work with the database provided by the Johns Hopkins University.
Financial support from the Swiss National Science Foundation (SNSF) under Grant No. 144645 is gratefully acknowledged.

\bibliography{biblio}

\begin{thebibliography}{49}%
\makeatletter
\providecommand \@ifxundefined [1]{%
 \@ifx{#1\undefined}
}%
\providecommand \@ifnum [1]{%
 \ifnum #1\expandafter \@firstoftwo
 \else \expandafter \@secondoftwo
 \fi
}%
\providecommand \@ifx [1]{%
 \ifx #1\expandafter \@firstoftwo
 \else \expandafter \@secondoftwo
 \fi
}%
\providecommand \natexlab [1]{#1}%
\providecommand \enquote  [1]{``#1''}%
\providecommand \bibnamefont  [1]{#1}%
\providecommand \bibfnamefont [1]{#1}%
\providecommand \citenamefont [1]{#1}%
\providecommand \href@noop [0]{\@secondoftwo}%
\providecommand \href [0]{\begingroup \@sanitize@url \@href}%
\providecommand \@href[1]{\@@startlink{#1}\@@href}%
\providecommand \@@href[1]{\endgroup#1\@@endlink}%
\providecommand \@sanitize@url [0]{\catcode `\\12\catcode `\$12\catcode
  `\&12\catcode `\#12\catcode `\^12\catcode `\_12\catcode `\%12\relax}%
\providecommand \@@startlink[1]{}%
\providecommand \@@endlink[0]{}%
\providecommand \url  [0]{\begingroup\@sanitize@url \@url }%
\providecommand \@url [1]{\endgroup\@href {#1}{\urlprefix }}%
\providecommand \urlprefix  [0]{URL }%
\providecommand \Eprint [0]{\href }%
\providecommand \doibase [0]{http://dx.doi.org/}%
\providecommand \selectlanguage [0]{\@gobble}%
\providecommand \bibinfo  [0]{\@secondoftwo}%
\providecommand \bibfield  [0]{\@secondoftwo}%
\providecommand \translation [1]{[#1]}%
\providecommand \BibitemOpen [0]{}%
\providecommand \bibitemStop [0]{}%
\providecommand \bibitemNoStop [0]{.\EOS\space}%
\providecommand \EOS [0]{\spacefactor3000\relax}%
\providecommand \BibitemShut  [1]{\csname bibitem#1\endcsname}%
\let\auto@bib@innerbib\@empty
\bibitem [{\citenamefont {Sirignano}(1999)}]{Engineering2}%
  \BibitemOpen
  \bibfield  {author} {\bibinfo {author} {\bibfnamefont {William~A}\
  \bibnamefont {Sirignano}},\ }\href@noop {} {\emph {\bibinfo {title} {Fluid
  dynamics and transport of droplets and sprays}}}\ (\bibinfo  {publisher}
  {Cambridge University Press},\ \bibinfo {year} {1999})\BibitemShut {NoStop}%
\bibitem [{\citenamefont {Lee}\ \emph {et~al.}(2002)\citenamefont {Lee},
  \citenamefont {Cole}, \citenamefont {Sekar}, \citenamefont {Choi},
  \citenamefont {Kang}, \citenamefont {Bae},\ and\ \citenamefont
  {Shin}}]{lee2002morphological}%
  \BibitemOpen
  \bibfield  {author} {\bibinfo {author} {\bibfnamefont {Kyeong~O}\
  \bibnamefont {Lee}}, \bibinfo {author} {\bibfnamefont {Roger}\ \bibnamefont
  {Cole}}, \bibinfo {author} {\bibfnamefont {Raj}\ \bibnamefont {Sekar}},
  \bibinfo {author} {\bibfnamefont {Mun~Y}\ \bibnamefont {Choi}}, \bibinfo
  {author} {\bibfnamefont {Jin~S}\ \bibnamefont {Kang}}, \bibinfo {author}
  {\bibfnamefont {Choong~S}\ \bibnamefont {Bae}}, \ and\ \bibinfo {author}
  {\bibfnamefont {Hyun~D}\ \bibnamefont {Shin}},\ }\bibfield  {title} {\enquote
  {\bibinfo {title} {Morphological investigation of the microstructure,
  dimensions, and fractal geometry of diesel particulates},}\ }\href@noop {}
  {\bibfield  {journal} {\bibinfo  {journal} {Proceedings of the Combustion
  Institute}\ }\textbf {\bibinfo {volume} {29}},\ \bibinfo {pages} {647--653}
  (\bibinfo {year} {2002})}\BibitemShut {NoStop}%
\bibitem [{\citenamefont {Crowe}\ \emph {et~al.}(1998)\citenamefont {Crowe},
  \citenamefont {Sommerfeld},\ and\ \citenamefont {Tsuji}}]{Engineering1}%
  \BibitemOpen
  \bibfield  {author} {\bibinfo {author} {\bibfnamefont {C}~\bibnamefont
  {Crowe}}, \bibinfo {author} {\bibfnamefont {M}~\bibnamefont {Sommerfeld}}, \
  and\ \bibinfo {author} {\bibfnamefont {Y}~\bibnamefont {Tsuji}},\ }\bibfield
  {title} {\enquote {\bibinfo {title} {Multiphase flows with droplets and
  particles crc press},}\ }\href@noop {} {\bibfield  {journal} {\bibinfo
  {journal} {Boca Raton, FL}\ } (\bibinfo {year} {1998})}\BibitemShut {NoStop}%
\bibitem [{\citenamefont {Hidy}(2012)}]{hidy2012aerosols}%
  \BibitemOpen
  \bibfield  {author} {\bibinfo {author} {\bibfnamefont {George}\ \bibnamefont
  {Hidy}},\ }\href@noop {} {\emph {\bibinfo {title} {Aerosols: an industrial
  and environmental science}}}\ (\bibinfo  {publisher} {Elsevier},\ \bibinfo
  {year} {2012})\BibitemShut {NoStop}%
\bibitem [{\citenamefont {Falkovich}\ \emph {et~al.}(2002)\citenamefont
  {Falkovich}, \citenamefont {Fouxon},\ and\ \citenamefont {Stepanov}}]{FFS}%
  \BibitemOpen
  \bibfield  {author} {\bibinfo {author} {\bibfnamefont {G}~\bibnamefont
  {Falkovich}}, \bibinfo {author} {\bibfnamefont {A}~\bibnamefont {Fouxon}}, \
  and\ \bibinfo {author} {\bibfnamefont {MG}~\bibnamefont {Stepanov}},\
  }\bibfield  {title} {\enquote {\bibinfo {title} {Acceleration of rain
  initiation by cloud turbulenceacceleration of rain initiation by cloud
  turbulence},}\ }\href@noop {} {\bibfield  {journal} {\bibinfo  {journal}
  {Nature}\ }\textbf {\bibinfo {volume} {419}},\ \bibinfo {pages} {151--154}
  (\bibinfo {year} {2002})}\BibitemShut {NoStop}%
\bibitem [{\citenamefont {Shaw}(2003)}]{shaw2003particle}%
  \BibitemOpen
  \bibfield  {author} {\bibinfo {author} {\bibfnamefont {Raymond~A}\
  \bibnamefont {Shaw}},\ }\bibfield  {title} {\enquote {\bibinfo {title}
  {Particle-turbulence interactions in atmospheric clouds},}\ }\href@noop {}
  {\bibfield  {journal} {\bibinfo  {journal} {Annual Review of Fluid
  Mechanics}\ }\textbf {\bibinfo {volume} {35}},\ \bibinfo {pages} {183--227}
  (\bibinfo {year} {2003})}\BibitemShut {NoStop}%
\bibitem [{\citenamefont {Williams}\ and\ \citenamefont
  {Crane}(1979)}]{williams1979drop}%
  \BibitemOpen
  \bibfield  {author} {\bibinfo {author} {\bibfnamefont {JJE}\ \bibnamefont
  {Williams}}\ and\ \bibinfo {author} {\bibfnamefont {RI}~\bibnamefont
  {Crane}},\ }\bibfield  {title} {\enquote {\bibinfo {title} {Drop coagulation
  in cross-over pipe flows of wet steam},}\ }\href@noop {} {\bibfield
  {journal} {\bibinfo  {journal} {Journal of Mechanical Engineering Science}\
  }\textbf {\bibinfo {volume} {21}},\ \bibinfo {pages} {357--360} (\bibinfo
  {year} {1979})}\BibitemShut {NoStop}%
\bibitem [{\citenamefont {Seinfeld}\ and\ \citenamefont
  {Pandis}(2012)}]{Seinfeld}%
  \BibitemOpen
  \bibfield  {author} {\bibinfo {author} {\bibfnamefont {John~H}\ \bibnamefont
  {Seinfeld}}\ and\ \bibinfo {author} {\bibfnamefont {Spyros~N}\ \bibnamefont
  {Pandis}},\ }\href@noop {} {\emph {\bibinfo {title} {Atmospheric chemistry
  and physics: from air pollution to climate change}}}\ (\bibinfo  {publisher}
  {John Wiley \& Sons},\ \bibinfo {year} {2012})\BibitemShut {NoStop}%
\bibitem [{\citenamefont {Reeks}(2014)}]{reeks2014transport}%
  \BibitemOpen
  \bibfield  {author} {\bibinfo {author} {\bibfnamefont {Michael~W}\
  \bibnamefont {Reeks}},\ }\bibfield  {title} {\enquote {\bibinfo {title}
  {Transport, mixing and agglomeration of particles in turbulent flows},}\ }in\
  \href@noop {} {\emph {\bibinfo {booktitle} {Journal of Physics: Conference
  Series}}},\ Vol.\ \bibinfo {volume} {530}\ (\bibinfo {organization} {IOP
  Publishing},\ \bibinfo {year} {2014})\ p.\ \bibinfo {pages}
  {012003}\BibitemShut {NoStop}%
\bibitem [{\citenamefont {Flagan}\ and\ \citenamefont
  {Seinfeld}(2013)}]{Flagan}%
  \BibitemOpen
  \bibfield  {author} {\bibinfo {author} {\bibfnamefont {Richard~C}\
  \bibnamefont {Flagan}}\ and\ \bibinfo {author} {\bibfnamefont {John~H}\
  \bibnamefont {Seinfeld}},\ }\href@noop {} {\emph {\bibinfo {title}
  {Fundamentals of air pollution engineering}}}\ (\bibinfo  {publisher}
  {Courier Corporation},\ \bibinfo {year} {2013})\BibitemShut {NoStop}%
\bibitem [{\citenamefont {Bec}\ \emph {et~al.}(2007{\natexlab{a}})\citenamefont
  {Bec}, \citenamefont {Cencini},\ and\ \citenamefont
  {Hillerbrand}}]{BecCenciniHillerbranddelta}%
  \BibitemOpen
  \bibfield  {author} {\bibinfo {author} {\bibfnamefont {J}~\bibnamefont
  {Bec}}, \bibinfo {author} {\bibfnamefont {M}~\bibnamefont {Cencini}}, \ and\
  \bibinfo {author} {\bibfnamefont {R}~\bibnamefont {Hillerbrand}},\ }\bibfield
   {title} {\enquote {\bibinfo {title} {Clustering of heavy particles in random
  self-similar flow},}\ }\href@noop {} {\bibfield  {journal} {\bibinfo
  {journal} {Phys. Rev. E}\ }\textbf {\bibinfo {volume} {75}},\ \bibinfo
  {pages} {025301} (\bibinfo {year} {2007}{\natexlab{a}})}\BibitemShut
  {NoStop}%
\bibitem [{\citenamefont {Sundaram}\ and\ \citenamefont
  {Collins}(1997)}]{sundaram1997collision}%
  \BibitemOpen
  \bibfield  {author} {\bibinfo {author} {\bibfnamefont {Shivshankar}\
  \bibnamefont {Sundaram}}\ and\ \bibinfo {author} {\bibfnamefont {Lance~R}\
  \bibnamefont {Collins}},\ }\bibfield  {title} {\enquote {\bibinfo {title}
  {Collision statistics in an isotropic particle-laden turbulent suspension.
  part 1. direct numerical simulations},}\ }\href@noop {} {\bibfield  {journal}
  {\bibinfo  {journal} {Journal of Fluid Mechanics}\ }\textbf {\bibinfo
  {volume} {335}},\ \bibinfo {pages} {75--109} (\bibinfo {year}
  {1997})}\BibitemShut {NoStop}%
\bibitem [{\citenamefont {Bec}\ \emph {et~al.}(2007{\natexlab{b}})\citenamefont
  {Bec}, \citenamefont {Biferale}, \citenamefont {Cencini}, \citenamefont
  {Lanotte}, \citenamefont {Musacchio},\ and\ \citenamefont
  {Toschi}}]{Stefano}%
  \BibitemOpen
  \bibfield  {author} {\bibinfo {author} {\bibfnamefont {Jeremie}\ \bibnamefont
  {Bec}}, \bibinfo {author} {\bibfnamefont {Luca}\ \bibnamefont {Biferale}},
  \bibinfo {author} {\bibfnamefont {Massimo}\ \bibnamefont {Cencini}}, \bibinfo
  {author} {\bibfnamefont {A}~\bibnamefont {Lanotte}}, \bibinfo {author}
  {\bibfnamefont {Stefano}\ \bibnamefont {Musacchio}}, \ and\ \bibinfo {author}
  {\bibfnamefont {Federico}\ \bibnamefont {Toschi}},\ }\bibfield  {title}
  {\enquote {\bibinfo {title} {Heavy particle concentration in turbulence at
  dissipative and inertial scales},}\ }\href@noop {} {\bibfield  {journal}
  {\bibinfo  {journal} {Phys. Rev. Lett.}\ }\textbf {\bibinfo {volume} {98}},\
  \bibinfo {pages} {084502} (\bibinfo {year} {2007}{\natexlab{b}})}\BibitemShut
  {NoStop}%
\bibitem [{\citenamefont {Jung}\ \emph {et~al.}(2008)\citenamefont {Jung},
  \citenamefont {Yeo},\ and\ \citenamefont {Lee}}]{JYL}%
  \BibitemOpen
  \bibfield  {author} {\bibinfo {author} {\bibfnamefont {Jaedal}\ \bibnamefont
  {Jung}}, \bibinfo {author} {\bibfnamefont {Kyongmin}\ \bibnamefont {Yeo}}, \
  and\ \bibinfo {author} {\bibfnamefont {Changhoon}\ \bibnamefont {Lee}},\
  }\bibfield  {title} {\enquote {\bibinfo {title} {Behavior of heavy particles
  in isotropic turbulence},}\ }\href@noop {} {\bibfield  {journal} {\bibinfo
  {journal} {Phys. Rev. E}\ }\textbf {\bibinfo {volume} {77}},\ \bibinfo
  {pages} {016307} (\bibinfo {year} {2008})}\BibitemShut {NoStop}%
\bibitem [{\citenamefont {Balkovsky}\ \emph {et~al.}(2001)\citenamefont
  {Balkovsky}, \citenamefont {Falkovich},\ and\ \citenamefont {Fouxon}}]{BFF}%
  \BibitemOpen
  \bibfield  {author} {\bibinfo {author} {\bibfnamefont {E}~\bibnamefont
  {Balkovsky}}, \bibinfo {author} {\bibfnamefont {Gregory}\ \bibnamefont
  {Falkovich}}, \ and\ \bibinfo {author} {\bibfnamefont {A}~\bibnamefont
  {Fouxon}},\ }\bibfield  {title} {\enquote {\bibinfo {title} {Intermittent
  distribution of inertial particles in turbulent flows},}\ }\href@noop {}
  {\bibfield  {journal} {\bibinfo  {journal} {Phys. Rev. Lett.}\ }\textbf
  {\bibinfo {volume} {86}},\ \bibinfo {pages} {2790} (\bibinfo {year}
  {2001})}\BibitemShut {NoStop}%
\bibitem [{\citenamefont {Cencini}\ \emph {et~al.}(2006)\citenamefont
  {Cencini}, \citenamefont {Bec}, \citenamefont {Biferale}, \citenamefont
  {Boffetta}, \citenamefont {Celani}, \citenamefont {Lanotte}, \citenamefont
  {Musacchio},\ and\ \citenamefont {Toschi}}]{CenciniBecBiferale}%
  \BibitemOpen
  \bibfield  {author} {\bibinfo {author} {\bibfnamefont {M}~\bibnamefont
  {Cencini}}, \bibinfo {author} {\bibfnamefont {J}~\bibnamefont {Bec}},
  \bibinfo {author} {\bibfnamefont {L.}~\bibnamefont {Biferale}}, \bibinfo
  {author} {\bibfnamefont {G.}~\bibnamefont {Boffetta}}, \bibinfo {author}
  {\bibfnamefont {A.}~\bibnamefont {Celani}}, \bibinfo {author} {\bibfnamefont
  {A.~S.}\ \bibnamefont {Lanotte}}, \bibinfo {author} {\bibfnamefont
  {S.}~\bibnamefont {Musacchio}}, \ and\ \bibinfo {author} {\bibfnamefont
  {F.}~\bibnamefont {Toschi}},\ }\bibfield  {title} {\enquote {\bibinfo {title}
  {Dynamics and statistics of heavy particles in turbulent flows},}\ }\href
  {\doibase 10.1080/14685240600675727} {\bibfield  {journal} {\bibinfo
  {journal} {Journal of Turbulence}\ }\textbf {\bibinfo {volume} {7}},\
  \bibinfo {pages} {N36} (\bibinfo {year} {2006})}\BibitemShut {NoStop}%
\bibitem [{\citenamefont {Calzavarini}\ \emph {et~al.}(2008)\citenamefont
  {Calzavarini}, \citenamefont {Cencini}, \citenamefont {Lohse},\ and\
  \citenamefont {Toschi}}]{Cencini}%
  \BibitemOpen
  \bibfield  {author} {\bibinfo {author} {\bibfnamefont {Enrico}\ \bibnamefont
  {Calzavarini}}, \bibinfo {author} {\bibfnamefont {Massimo}\ \bibnamefont
  {Cencini}}, \bibinfo {author} {\bibfnamefont {Detlef}\ \bibnamefont {Lohse}},
  \ and\ \bibinfo {author} {\bibfnamefont {Federico}\ \bibnamefont {Toschi}},\
  }\bibfield  {title} {\enquote {\bibinfo {title} {Quantifying
  turbulence-induced segregation of inertial particles},}\ }\href@noop {}
  {\bibfield  {journal} {\bibinfo  {journal} {Phys. Rev. Lett.}\ }\textbf
  {\bibinfo {volume} {101}},\ \bibinfo {pages} {084504} (\bibinfo {year}
  {2008})}\BibitemShut {NoStop}%
\bibitem [{\citenamefont {Monchaux}\ \emph {et~al.}(2010)\citenamefont
  {Monchaux}, \citenamefont {Bourgoin},\ and\ \citenamefont
  {Cartellier}}]{monchaux2010preferential}%
  \BibitemOpen
  \bibfield  {author} {\bibinfo {author} {\bibfnamefont {Romain}\ \bibnamefont
  {Monchaux}}, \bibinfo {author} {\bibfnamefont {Micka{\"e}l}\ \bibnamefont
  {Bourgoin}}, \ and\ \bibinfo {author} {\bibfnamefont {Alain}\ \bibnamefont
  {Cartellier}},\ }\bibfield  {title} {\enquote {\bibinfo {title} {Preferential
  concentration of heavy particles: a voronoi analysis},}\ }\href@noop {}
  {\bibfield  {journal} {\bibinfo  {journal} {Physics of Fluids
  (1994-present)}\ }\textbf {\bibinfo {volume} {22}},\ \bibinfo {pages}
  {103304} (\bibinfo {year} {2010})}\BibitemShut {NoStop}%
\bibitem [{\citenamefont {Eaton}\ and\ \citenamefont
  {Fessler}(1994)}]{eatonfessler}%
  \BibitemOpen
  \bibfield  {author} {\bibinfo {author} {\bibfnamefont {J.K.}\ \bibnamefont
  {Eaton}}\ and\ \bibinfo {author} {\bibfnamefont {J.R.}\ \bibnamefont
  {Fessler}},\ }\bibfield  {title} {\enquote {\bibinfo {title} {Preferential
  concentration of particles by turbulence},}\ }\href {\doibase
  http://dx.doi.org/10.1016/0301-9322(94)90072-8} {\bibfield  {journal}
  {\bibinfo  {journal} {International Journal of Multiphase Flow}\ }\textbf
  {\bibinfo {volume} {20}},\ \bibinfo {pages} {169 -- 209} (\bibinfo {year}
  {1994})}\BibitemShut {NoStop}%
\bibitem [{\citenamefont {Caporaloni}\ \emph {et~al.}(1975)\citenamefont
  {Caporaloni}, \citenamefont {Tampieri}, \citenamefont {Trombetti},\ and\
  \citenamefont {Vittori}}]{caporaloni}%
  \BibitemOpen
  \bibfield  {author} {\bibinfo {author} {\bibfnamefont {M}~\bibnamefont
  {Caporaloni}}, \bibinfo {author} {\bibfnamefont {F}~\bibnamefont {Tampieri}},
  \bibinfo {author} {\bibfnamefont {F}~\bibnamefont {Trombetti}}, \ and\
  \bibinfo {author} {\bibfnamefont {O}~\bibnamefont {Vittori}},\ }\bibfield
  {title} {\enquote {\bibinfo {title} {Transfer of particles in nonisotropic
  air turbulence},}\ }\href@noop {} {\bibfield  {journal} {\bibinfo  {journal}
  {Journal of the atmospheric sciences}\ }\textbf {\bibinfo {volume} {32}},\
  \bibinfo {pages} {565--568} (\bibinfo {year} {1975})}\BibitemShut {NoStop}%
\bibitem [{\citenamefont {Reeks}(1983)}]{reeksturbophoresis}%
  \BibitemOpen
  \bibfield  {author} {\bibinfo {author} {\bibfnamefont {MW}~\bibnamefont
  {Reeks}},\ }\bibfield  {title} {\enquote {\bibinfo {title} {The transport of
  discrete particles in inhomogeneous turbulence},}\ }\href@noop {} {\bibfield
  {journal} {\bibinfo  {journal} {Journal of aerosol science}\ }\textbf
  {\bibinfo {volume} {14}},\ \bibinfo {pages} {729--739} (\bibinfo {year}
  {1983})}\BibitemShut {NoStop}%
\bibitem [{\citenamefont {Young}\ and\ \citenamefont {Leeming}(1997)}]{young}%
  \BibitemOpen
  \bibfield  {author} {\bibinfo {author} {\bibfnamefont {John}\ \bibnamefont
  {Young}}\ and\ \bibinfo {author} {\bibfnamefont {Angus}\ \bibnamefont
  {Leeming}},\ }\bibfield  {title} {\enquote {\bibinfo {title} {A theory of
  particle deposition in turbulent pipe flow},}\ }\href@noop {} {\bibfield
  {journal} {\bibinfo  {journal} {Journal of Fluid Mechanics}\ }\textbf
  {\bibinfo {volume} {340}},\ \bibinfo {pages} {129--159} (\bibinfo {year}
  {1997})}\BibitemShut {NoStop}%
\bibitem [{\citenamefont {Kaftori}\ \emph
  {et~al.}(1995{\natexlab{a}})\citenamefont {Kaftori}, \citenamefont
  {Hetsroni},\ and\ \citenamefont {Banerjee}}]{kaftoria}%
  \BibitemOpen
  \bibfield  {author} {\bibinfo {author} {\bibfnamefont {D}~\bibnamefont
  {Kaftori}}, \bibinfo {author} {\bibfnamefont {G}~\bibnamefont {Hetsroni}}, \
  and\ \bibinfo {author} {\bibfnamefont {S}~\bibnamefont {Banerjee}},\
  }\bibfield  {title} {\enquote {\bibinfo {title} {Particle behavior in the
  turbulent boundary layer. i. motion, deposition, and entrainment},}\
  }\href@noop {} {\bibfield  {journal} {\bibinfo  {journal} {Physics of Fluids
  (1994-present)}\ }\textbf {\bibinfo {volume} {7}},\ \bibinfo {pages}
  {1095--1106} (\bibinfo {year} {1995}{\natexlab{a}})}\BibitemShut {NoStop}%
\bibitem [{\citenamefont {Kaftori}\ \emph
  {et~al.}(1995{\natexlab{b}})\citenamefont {Kaftori}, \citenamefont
  {Hetsroni},\ and\ \citenamefont {Banerjee}}]{kaftorib}%
  \BibitemOpen
  \bibfield  {author} {\bibinfo {author} {\bibfnamefont {D}~\bibnamefont
  {Kaftori}}, \bibinfo {author} {\bibfnamefont {G}~\bibnamefont {Hetsroni}}, \
  and\ \bibinfo {author} {\bibfnamefont {S}~\bibnamefont {Banerjee}},\
  }\bibfield  {title} {\enquote {\bibinfo {title} {Particle behavior in the
  turbulent boundary layer. ii. velocity and distribution profiles},}\
  }\href@noop {} {\bibfield  {journal} {\bibinfo  {journal} {Physics of Fluids
  (1994-present)}\ }\textbf {\bibinfo {volume} {7}},\ \bibinfo {pages}
  {1107--1121} (\bibinfo {year} {1995}{\natexlab{b}})}\BibitemShut {NoStop}%
\bibitem [{\citenamefont {Righetti}\ and\ \citenamefont
  {Romano}(2004)}]{righetti}%
  \BibitemOpen
  \bibfield  {author} {\bibinfo {author} {\bibfnamefont {M}~\bibnamefont
  {Righetti}}\ and\ \bibinfo {author} {\bibfnamefont {Giovanni~Paolo}\
  \bibnamefont {Romano}},\ }\bibfield  {title} {\enquote {\bibinfo {title}
  {Particle--fluid interactions in a plane near-wall turbulent flow},}\
  }\href@noop {} {\bibfield  {journal} {\bibinfo  {journal} {Journal of Fluid
  Mechanics}\ }\textbf {\bibinfo {volume} {505}},\ \bibinfo {pages} {93--121}
  (\bibinfo {year} {2004})}\BibitemShut {NoStop}%
\bibitem [{\citenamefont {Marchioli}\ and\ \citenamefont
  {Soldati}(2002)}]{marchioli}%
  \BibitemOpen
  \bibfield  {author} {\bibinfo {author} {\bibfnamefont {Cristian}\
  \bibnamefont {Marchioli}}\ and\ \bibinfo {author} {\bibfnamefont {Alfredo}\
  \bibnamefont {Soldati}},\ }\bibfield  {title} {\enquote {\bibinfo {title}
  {Mechanisms for particle transfer and segregation in a turbulent boundary
  layer},}\ }\href@noop {} {\bibfield  {journal} {\bibinfo  {journal} {Journal
  of fluid Mechanics}\ }\textbf {\bibinfo {volume} {468}},\ \bibinfo {pages}
  {283--315} (\bibinfo {year} {2002})}\BibitemShut {NoStop}%
\bibitem [{\citenamefont {Picciotto}\ \emph {et~al.}(2005)\citenamefont
  {Picciotto}, \citenamefont {Marchioli},\ and\ \citenamefont
  {Soldati}}]{picciotto}%
  \BibitemOpen
  \bibfield  {author} {\bibinfo {author} {\bibfnamefont {Maurizio}\
  \bibnamefont {Picciotto}}, \bibinfo {author} {\bibfnamefont {Cristian}\
  \bibnamefont {Marchioli}}, \ and\ \bibinfo {author} {\bibfnamefont {Alfredo}\
  \bibnamefont {Soldati}},\ }\bibfield  {title} {\enquote {\bibinfo {title}
  {Characterization of near-wall accumulation regions for inertial particles in
  turbulent boundary layers},}\ }\href@noop {} {\bibfield  {journal} {\bibinfo
  {journal} {Physics of Fluids (1994-present)}\ }\textbf {\bibinfo {volume}
  {17}},\ \bibinfo {pages} {098101} (\bibinfo {year} {2005})}\BibitemShut
  {NoStop}%
\bibitem [{\citenamefont {Picano}\ \emph {et~al.}(2009)\citenamefont {Picano},
  \citenamefont {Sardina},\ and\ \citenamefont {Casciola}}]{picano}%
  \BibitemOpen
  \bibfield  {author} {\bibinfo {author} {\bibfnamefont {F}~\bibnamefont
  {Picano}}, \bibinfo {author} {\bibfnamefont {G}~\bibnamefont {Sardina}}, \
  and\ \bibinfo {author} {\bibfnamefont {CM}~\bibnamefont {Casciola}},\
  }\bibfield  {title} {\enquote {\bibinfo {title} {Spatial development of
  particle-laden turbulent pipe flow},}\ }\href@noop {} {\bibfield  {journal}
  {\bibinfo  {journal} {Physics of Fluids (1994-present)}\ }\textbf {\bibinfo
  {volume} {21}},\ \bibinfo {pages} {093305} (\bibinfo {year}
  {2009})}\BibitemShut {NoStop}%
\bibitem [{\citenamefont {Sardina}\ \emph {et~al.}(2012)\citenamefont
  {Sardina}, \citenamefont {Schlatter}, \citenamefont {Brandt}, \citenamefont
  {Picano},\ and\ \citenamefont {Casciola}}]{sardina}%
  \BibitemOpen
  \bibfield  {author} {\bibinfo {author} {\bibfnamefont {G}~\bibnamefont
  {Sardina}}, \bibinfo {author} {\bibfnamefont {Philipp}\ \bibnamefont
  {Schlatter}}, \bibinfo {author} {\bibfnamefont {Luca}\ \bibnamefont
  {Brandt}}, \bibinfo {author} {\bibfnamefont {F}~\bibnamefont {Picano}}, \
  and\ \bibinfo {author} {\bibfnamefont {Carlo~Massimo}\ \bibnamefont
  {Casciola}},\ }\bibfield  {title} {\enquote {\bibinfo {title} {Wall
  accumulation and spatial localization in particle-laden wall flows},}\
  }\href@noop {} {\bibfield  {journal} {\bibinfo  {journal} {Journal of Fluid
  Mechanics}\ }\textbf {\bibinfo {volume} {699}},\ \bibinfo {pages} {50--78}
  (\bibinfo {year} {2012})}\BibitemShut {NoStop}%
\bibitem [{\citenamefont {De~Lillo}\ \emph {et~al.}(2016)\citenamefont
  {De~Lillo}, \citenamefont {Cencini}, \citenamefont {Musacchio},\ and\
  \citenamefont {Boffetta}}]{de2016clustering}%
  \BibitemOpen
  \bibfield  {author} {\bibinfo {author} {\bibfnamefont {Filippo}\ \bibnamefont
  {De~Lillo}}, \bibinfo {author} {\bibfnamefont {Massimo}\ \bibnamefont
  {Cencini}}, \bibinfo {author} {\bibfnamefont {Stefano}\ \bibnamefont
  {Musacchio}}, \ and\ \bibinfo {author} {\bibfnamefont {Guido}\ \bibnamefont
  {Boffetta}},\ }\bibfield  {title} {\enquote {\bibinfo {title} {Clustering and
  turbophoresis in a shear flow without walls},}\ }\href@noop {} {\bibfield
  {journal} {\bibinfo  {journal} {Physics of Fluids (1994-present)}\ }\textbf
  {\bibinfo {volume} {28}},\ \bibinfo {pages} {035104} (\bibinfo {year}
  {2016})}\BibitemShut {NoStop}%
\bibitem [{\citenamefont {Schmidt}\ \emph {et~al.}(2016)\citenamefont
  {Schmidt}, \citenamefont {Fouxon}, \citenamefont {Krug}, \citenamefont {van
  Reeuwijk},\ and\ \citenamefont {Holzner}}]{Schmidt}%
  \BibitemOpen
  \bibfield  {author} {\bibinfo {author} {\bibfnamefont {Lukas}\ \bibnamefont
  {Schmidt}}, \bibinfo {author} {\bibfnamefont {Itzhak}\ \bibnamefont
  {Fouxon}}, \bibinfo {author} {\bibfnamefont {Dominik}\ \bibnamefont {Krug}},
  \bibinfo {author} {\bibfnamefont {Maarten}\ \bibnamefont {van Reeuwijk}}, \
  and\ \bibinfo {author} {\bibfnamefont {Markus}\ \bibnamefont {Holzner}},\
  }\bibfield  {title} {\enquote {\bibinfo {title} {Clustering of particles in
  turbulence due to phoresis},}\ }\href {\doibase 10.1103/PhysRevE.93.063110}
  {\bibfield  {journal} {\bibinfo  {journal} {Phys. Rev. E}\ }\textbf {\bibinfo
  {volume} {93}},\ \bibinfo {pages} {063110} (\bibinfo {year}
  {2016})}\BibitemShut {NoStop}%
\bibitem [{\citenamefont {Fouxon}(2012)}]{fouxon1}%
  \BibitemOpen
  \bibfield  {author} {\bibinfo {author} {\bibfnamefont {Itzhak}\ \bibnamefont
  {Fouxon}},\ }\bibfield  {title} {\enquote {\bibinfo {title} {Distribution of
  particles and bubbles in turbulence at a small stokes number},}\ }\href@noop
  {} {\bibfield  {journal} {\bibinfo  {journal} {Phys. Rev. Lett.}\ }\textbf
  {\bibinfo {volume} {108}},\ \bibinfo {pages} {134502} (\bibinfo {year}
  {2012})}\BibitemShut {NoStop}%
\bibitem [{\citenamefont {Fouxon}\ \emph {et~al.}(2015)\citenamefont {Fouxon},
  \citenamefont {Park}, \citenamefont {Harduf},\ and\ \citenamefont
  {Lee}}]{fphl}%
  \BibitemOpen
  \bibfield  {author} {\bibinfo {author} {\bibfnamefont {Itzhak}\ \bibnamefont
  {Fouxon}}, \bibinfo {author} {\bibfnamefont {Yongnam}\ \bibnamefont {Park}},
  \bibinfo {author} {\bibfnamefont {Roei}\ \bibnamefont {Harduf}}, \ and\
  \bibinfo {author} {\bibfnamefont {Changhoon}\ \bibnamefont {Lee}},\
  }\bibfield  {title} {\enquote {\bibinfo {title} {Inhomogeneous distribution
  of water droplets in cloud turbulence},}\ }\href@noop {} {\bibfield
  {journal} {\bibinfo  {journal} {Phys. Rev. E}\ }\textbf {\bibinfo {volume}
  {92}},\ \bibinfo {pages} {033001} (\bibinfo {year} {2015})}\BibitemShut
  {NoStop}%
\bibitem [{\citenamefont {Maxey}\ and\ \citenamefont
  {Riley}(1983)}]{MaxeyRiley}%
  \BibitemOpen
  \bibfield  {author} {\bibinfo {author} {\bibfnamefont {Martin~R}\
  \bibnamefont {Maxey}}\ and\ \bibinfo {author} {\bibfnamefont {James~J}\
  \bibnamefont {Riley}},\ }\bibfield  {title} {\enquote {\bibinfo {title}
  {Equation of motion for a small rigid sphere in a nonuniform flow},}\
  }\href@noop {} {\bibfield  {journal} {\bibinfo  {journal} {Physics of Fluids
  (1958-1988)}\ }\textbf {\bibinfo {volume} {26}},\ \bibinfo {pages} {883--889}
  (\bibinfo {year} {1983})}\BibitemShut {NoStop}%
\bibitem [{\citenamefont {Maxey}(1987)}]{Maxey}%
  \BibitemOpen
  \bibfield  {author} {\bibinfo {author} {\bibfnamefont {MR}~\bibnamefont
  {Maxey}},\ }\bibfield  {title} {\enquote {\bibinfo {title} {The gravitational
  settling of aerosol particles in homogeneous turbulence and random flow
  fields},}\ }\href@noop {} {\bibfield  {journal} {\bibinfo  {journal} {J.
  Fluid. Mech.}\ }\textbf {\bibinfo {volume} {174}},\ \bibinfo {pages}
  {441--465} (\bibinfo {year} {1987})}\BibitemShut {NoStop}%
\bibitem [{\citenamefont {Frisch}(1995)}]{frisch}%
  \BibitemOpen
  \bibfield  {author} {\bibinfo {author} {\bibfnamefont {Uriel}\ \bibnamefont
  {Frisch}},\ }\href@noop {} {\emph {\bibinfo {title} {Turbulence: the legacy
  of AN Kolmogorov}}}\ (\bibinfo  {publisher} {Cambridge University Press},\
  \bibinfo {year} {1995})\BibitemShut {NoStop}%
\bibitem [{\citenamefont {Batchelor}(1959)}]{Batchelor}%
  \BibitemOpen
  \bibfield  {author} {\bibinfo {author} {\bibfnamefont {GK}~\bibnamefont
  {Batchelor}},\ }\bibfield  {title} {\enquote {\bibinfo {title} {Small-scale
  variation of convected quantities like temperature in turbulent fluid part 1.
  general discussion and the case of small conductivity},}\ }\href@noop {}
  {\bibfield  {journal} {\bibinfo  {journal} {J. Fluid. Mech.}\ }\textbf
  {\bibinfo {volume} {5}},\ \bibinfo {pages} {113--133} (\bibinfo {year}
  {1959})}\BibitemShut {NoStop}%
\bibitem [{\citenamefont {Sinai}(1972)}]{Sinai}%
  \BibitemOpen
  \bibfield  {author} {\bibinfo {author} {\bibfnamefont {Yakov~G}\ \bibnamefont
  {Sinai}},\ }\bibfield  {title} {\enquote {\bibinfo {title} {Gibbs measures in
  ergodic theory},}\ }\href@noop {} {\bibfield  {journal} {\bibinfo  {journal}
  {Russian Mathematical Surveys}\ }\textbf {\bibinfo {volume} {27}},\ \bibinfo
  {pages} {21} (\bibinfo {year} {1972})}\BibitemShut {NoStop}%
\bibitem [{\citenamefont {Devenish}\ \emph {et~al.}(2012)\citenamefont
  {Devenish}, \citenamefont {Bartello}, \citenamefont {Brenguier},
  \citenamefont {Collins}, \citenamefont {Grabowski}, \citenamefont
  {IJzermans}, \citenamefont {Malinowski}, \citenamefont {Reeks}, \citenamefont
  {Vassilicos}, \citenamefont {Wang} \emph {et~al.}}]{review}%
  \BibitemOpen
  \bibfield  {author} {\bibinfo {author} {\bibfnamefont {BJ}~\bibnamefont
  {Devenish}}, \bibinfo {author} {\bibfnamefont {P}~\bibnamefont {Bartello}},
  \bibinfo {author} {\bibfnamefont {J-L}\ \bibnamefont {Brenguier}}, \bibinfo
  {author} {\bibfnamefont {LR}~\bibnamefont {Collins}}, \bibinfo {author}
  {\bibfnamefont {WW}~\bibnamefont {Grabowski}}, \bibinfo {author}
  {\bibfnamefont {RHA}\ \bibnamefont {IJzermans}}, \bibinfo {author}
  {\bibfnamefont {SP}~\bibnamefont {Malinowski}}, \bibinfo {author}
  {\bibfnamefont {MW}~\bibnamefont {Reeks}}, \bibinfo {author} {\bibfnamefont
  {JC}~\bibnamefont {Vassilicos}}, \bibinfo {author} {\bibfnamefont {L-P}\
  \bibnamefont {Wang}},  \emph {et~al.},\ }\bibfield  {title} {\enquote
  {\bibinfo {title} {Droplet growth in warm turbulent clouds},}\ }\href@noop {}
  {\bibfield  {journal} {\bibinfo  {journal} {Q. J. Roy. Meteor. Soc.}\
  }\textbf {\bibinfo {volume} {138}},\ \bibinfo {pages} {1401--1429} (\bibinfo
  {year} {2012})}\BibitemShut {NoStop}%
\bibitem [{\citenamefont {Falkovich}\ and\ \citenamefont {Fouxon}(2004)}]{ff}%
  \BibitemOpen
  \bibfield  {author} {\bibinfo {author} {\bibfnamefont {Gregory}\ \bibnamefont
  {Falkovich}}\ and\ \bibinfo {author} {\bibfnamefont {Alexander}\ \bibnamefont
  {Fouxon}},\ }\bibfield  {title} {\enquote {\bibinfo {title} {Entropy
  production and extraction in dynamical systems and turbulence},}\ }\href@noop
  {} {\bibfield  {journal} {\bibinfo  {journal} {New. J. Phys.}\ }\textbf
  {\bibinfo {volume} {6}},\ \bibinfo {pages} {50} (\bibinfo {year}
  {2004})}\BibitemShut {NoStop}%
\bibitem [{\citenamefont {Kaplan}\ and\ \citenamefont {Yorke}(1979)}]{ky}%
  \BibitemOpen
  \bibfield  {author} {\bibinfo {author} {\bibfnamefont {JamesL.}\ \bibnamefont
  {Kaplan}}\ and\ \bibinfo {author} {\bibfnamefont {JamesA.}\ \bibnamefont
  {Yorke}},\ }\bibfield  {title} {\enquote {\bibinfo {title} {Chaotic behavior
  of multidimensional difference equations},}\ }in\ \href {\doibase
  10.1007/BFb0064319} {\emph {\bibinfo {booktitle} {Functional Differential
  Equations and Approximation of Fixed Points}}},\ \bibinfo {series} {Lecture
  Notes in Mathematics}, Vol.\ \bibinfo {volume} {730},\ \bibinfo {editor}
  {edited by\ \bibinfo {editor} {\bibfnamefont {Heinz-Otto}\ \bibnamefont
  {Peitgen}}\ and\ \bibinfo {editor} {\bibfnamefont {Hans-Otto}\ \bibnamefont
  {Walther}}}\ (\bibinfo  {publisher} {Springer Berlin Heidelberg},\ \bibinfo
  {year} {1979})\ pp.\ \bibinfo {pages} {204--227}\BibitemShut {NoStop}%
\bibitem [{\citenamefont {Fouxon}(2011)}]{fouxonnlin}%
  \BibitemOpen
  \bibfield  {author} {\bibinfo {author} {\bibfnamefont {Itzhak}\ \bibnamefont
  {Fouxon}},\ }\bibfield  {title} {\enquote {\bibinfo {title} {Evolution to a
  singular measure and two sums of lyapunov exponents},}\ }\href@noop {}
  {\bibfield  {journal} {\bibinfo  {journal} {Journal of Statistical Mechanics:
  Theory and Experiment}\ }\textbf {\bibinfo {volume} {2011}},\ \bibinfo
  {pages} {L02001} (\bibinfo {year} {2011})}\BibitemShut {NoStop}%
\bibitem [{\citenamefont {Perlman}\ \emph {et~al.}(2007)\citenamefont
  {Perlman}, \citenamefont {Burns}, \citenamefont {Li},\ and\ \citenamefont
  {Meneveau}}]{perlman2007data}%
  \BibitemOpen
  \bibfield  {author} {\bibinfo {author} {\bibfnamefont {Eric}\ \bibnamefont
  {Perlman}}, \bibinfo {author} {\bibfnamefont {Randal}\ \bibnamefont {Burns}},
  \bibinfo {author} {\bibfnamefont {Yi}~\bibnamefont {Li}}, \ and\ \bibinfo
  {author} {\bibfnamefont {Charles}\ \bibnamefont {Meneveau}},\ }\bibfield
  {title} {\enquote {\bibinfo {title} {Data exploration of turbulence
  simulations using a database cluster},}\ }in\ \href@noop {} {\emph {\bibinfo
  {booktitle} {Proceedings of the 2007 ACM/IEEE conference on
  Supercomputing}}}\ (\bibinfo {organization} {ACM},\ \bibinfo {year} {2007})\
  p.~\bibinfo {pages} {23}\BibitemShut {NoStop}%
\bibitem [{\citenamefont {Li}\ \emph {et~al.}(2008)\citenamefont {Li},
  \citenamefont {Perlman}, \citenamefont {Wan}, \citenamefont {Yang},
  \citenamefont {Meneveau}, \citenamefont {Burns}, \citenamefont {Chen},
  \citenamefont {Szalay},\ and\ \citenamefont {Eyink}}]{li2008public}%
  \BibitemOpen
  \bibfield  {author} {\bibinfo {author} {\bibfnamefont {Yi}~\bibnamefont
  {Li}}, \bibinfo {author} {\bibfnamefont {Eric}\ \bibnamefont {Perlman}},
  \bibinfo {author} {\bibfnamefont {Minping}\ \bibnamefont {Wan}}, \bibinfo
  {author} {\bibfnamefont {Yunke}\ \bibnamefont {Yang}}, \bibinfo {author}
  {\bibfnamefont {Charles}\ \bibnamefont {Meneveau}}, \bibinfo {author}
  {\bibfnamefont {Randal}\ \bibnamefont {Burns}}, \bibinfo {author}
  {\bibfnamefont {Shiyi}\ \bibnamefont {Chen}}, \bibinfo {author}
  {\bibfnamefont {Alexander}\ \bibnamefont {Szalay}}, \ and\ \bibinfo {author}
  {\bibfnamefont {Gregory}\ \bibnamefont {Eyink}},\ }\bibfield  {title}
  {\enquote {\bibinfo {title} {A public turbulence database cluster and
  applications to study lagrangian evolution of velocity increments in
  turbulence},}\ }\href@noop {} {\bibfield  {journal} {\bibinfo  {journal}
  {Journal of Turbulence}\ ,\ \bibinfo {pages} {N31}} (\bibinfo {year}
  {2008})}\BibitemShut {NoStop}%
\bibitem [{\citenamefont {Yu}\ \emph {et~al.}(2012)\citenamefont {Yu},
  \citenamefont {Kanov}, \citenamefont {Perlman}, \citenamefont {Graham},
  \citenamefont {Frederix}, \citenamefont {Burns}, \citenamefont {Szalay},
  \citenamefont {Eyink},\ and\ \citenamefont {Meneveau}}]{yu2012studying}%
  \BibitemOpen
  \bibfield  {author} {\bibinfo {author} {\bibfnamefont {Huidan}\ \bibnamefont
  {Yu}}, \bibinfo {author} {\bibfnamefont {Kalin}\ \bibnamefont {Kanov}},
  \bibinfo {author} {\bibfnamefont {Eric}\ \bibnamefont {Perlman}}, \bibinfo
  {author} {\bibfnamefont {Jason}\ \bibnamefont {Graham}}, \bibinfo {author}
  {\bibfnamefont {Edo}\ \bibnamefont {Frederix}}, \bibinfo {author}
  {\bibfnamefont {Randal}\ \bibnamefont {Burns}}, \bibinfo {author}
  {\bibfnamefont {Alexander}\ \bibnamefont {Szalay}}, \bibinfo {author}
  {\bibfnamefont {Gregory}\ \bibnamefont {Eyink}}, \ and\ \bibinfo {author}
  {\bibfnamefont {Charles}\ \bibnamefont {Meneveau}},\ }\bibfield  {title}
  {\enquote {\bibinfo {title} {Studying lagrangian dynamics of turbulence using
  on-demand fluid particle tracking in a public turbulence database},}\
  }\href@noop {} {\bibfield  {journal} {\bibinfo  {journal} {Journal of
  Turbulence}\ ,\ \bibinfo {pages} {N12}} (\bibinfo {year} {2012})}\BibitemShut
  {NoStop}%
\bibitem [{\citenamefont {Graham}\ \emph {et~al.}(2016)\citenamefont {Graham},
  \citenamefont {Kanov}, \citenamefont {Yang}, \citenamefont {Lee},
  \citenamefont {Malaya}, \citenamefont {Lalescu}, \citenamefont {Burns},
  \citenamefont {Eyink}, \citenamefont {Szalay}, \citenamefont {Moser} \emph
  {et~al.}}]{graham2016web}%
  \BibitemOpen
  \bibfield  {author} {\bibinfo {author} {\bibfnamefont {J}~\bibnamefont
  {Graham}}, \bibinfo {author} {\bibfnamefont {K}~\bibnamefont {Kanov}},
  \bibinfo {author} {\bibfnamefont {XIA}\ \bibnamefont {Yang}}, \bibinfo
  {author} {\bibfnamefont {M}~\bibnamefont {Lee}}, \bibinfo {author}
  {\bibfnamefont {N}~\bibnamefont {Malaya}}, \bibinfo {author} {\bibfnamefont
  {CC}~\bibnamefont {Lalescu}}, \bibinfo {author} {\bibfnamefont
  {R}~\bibnamefont {Burns}}, \bibinfo {author} {\bibfnamefont {G}~\bibnamefont
  {Eyink}}, \bibinfo {author} {\bibfnamefont {A}~\bibnamefont {Szalay}},
  \bibinfo {author} {\bibfnamefont {RD}~\bibnamefont {Moser}},  \emph
  {et~al.},\ }\bibfield  {title} {\enquote {\bibinfo {title} {A web services
  accessible database of turbulent channel flow and its use for testing a new
  integral wall model for les},}\ }\href@noop {} {\bibfield  {journal}
  {\bibinfo  {journal} {Journal of Turbulence}\ }\textbf {\bibinfo {volume}
  {17}},\ \bibinfo {pages} {181--215} (\bibinfo {year} {2016})}\BibitemShut
  {NoStop}%
\bibitem [{\citenamefont {Meneveau}\ \emph {et~al.}(2016)\citenamefont
  {Meneveau}, \citenamefont {Johnson}, \citenamefont {Hamilton},\ and\
  \citenamefont {Burns}}]{meneveau2016analysis}%
  \BibitemOpen
  \bibfield  {author} {\bibinfo {author} {\bibfnamefont {Charles}\ \bibnamefont
  {Meneveau}}, \bibinfo {author} {\bibfnamefont {Perry}\ \bibnamefont
  {Johnson}}, \bibinfo {author} {\bibfnamefont {Stephen}\ \bibnamefont
  {Hamilton}}, \ and\ \bibinfo {author} {\bibfnamefont {Randal}\ \bibnamefont
  {Burns}},\ }\bibfield  {title} {\enquote {\bibinfo {title} {Analysis of
  lagrangian stretching in turbulent channel flow using a database
  task-parallel particle tracking approach},}\ }\href@noop {} {\bibfield
  {journal} {\bibinfo  {journal} {Bulletin of the American Physical Society}\
  }\textbf {\bibinfo {volume} {61}} (\bibinfo {year} {2016})}\BibitemShut
  {NoStop}%
\bibitem [{\citenamefont {Saw}\ \emph {et~al.}(2012)\citenamefont {Saw},
  \citenamefont {Salazar}, \citenamefont {Collins},\ and\ \citenamefont
  {Shaw}}]{saw1}%
  \BibitemOpen
  \bibfield  {author} {\bibinfo {author} {\bibfnamefont {Ewe-Wei}\ \bibnamefont
  {Saw}}, \bibinfo {author} {\bibfnamefont {Juan~PLC}\ \bibnamefont {Salazar}},
  \bibinfo {author} {\bibfnamefont {Lance~R}\ \bibnamefont {Collins}}, \ and\
  \bibinfo {author} {\bibfnamefont {Raymond~A}\ \bibnamefont {Shaw}},\
  }\bibfield  {title} {\enquote {\bibinfo {title} {Spatial clustering of
  polydisperse inertial particles in turbulence: I. comparing simulation with
  theory},}\ }\href@noop {} {\bibfield  {journal} {\bibinfo  {journal} {New. J.
  Phys.}\ }\textbf {\bibinfo {volume} {14}},\ \bibinfo {pages} {105030}
  (\bibinfo {year} {2012})}\BibitemShut {NoStop}%
\bibitem [{\citenamefont {Chun}\ \emph {et~al.}(2005)\citenamefont {Chun},
  \citenamefont {Koch}, \citenamefont {Rani}, \citenamefont {Ahluwalia},\ and\
  \citenamefont {Collins}}]{Collinstwo}%
  \BibitemOpen
  \bibfield  {author} {\bibinfo {author} {\bibfnamefont {Jaehun}\ \bibnamefont
  {Chun}}, \bibinfo {author} {\bibfnamefont {Donald~L}\ \bibnamefont {Koch}},
  \bibinfo {author} {\bibfnamefont {Sarma~L}\ \bibnamefont {Rani}}, \bibinfo
  {author} {\bibfnamefont {Aruj}\ \bibnamefont {Ahluwalia}}, \ and\ \bibinfo
  {author} {\bibfnamefont {Lance~R}\ \bibnamefont {Collins}},\ }\bibfield
  {title} {\enquote {\bibinfo {title} {Clustering of aerosol particles in
  isotropic turbulence},}\ }\href@noop {} {\bibfield  {journal} {\bibinfo
  {journal} {J. Fluid Mech.}\ }\textbf {\bibinfo {volume} {536}},\ \bibinfo
  {pages} {219--251} (\bibinfo {year} {2005})}\BibitemShut {NoStop}%
\end{thebibliography}%

\end{document}